\documentclass[JHEP,12pt]{article}

\usepackage{graphics,color,soul}
\usepackage{graphicx}
\usepackage{amssymb}
\usepackage{jheppub}
\usepackage{lmodern,mathtools}
\usepackage{caption}
\usepackage{subcaption}
\usepackage{booktabs}
\usepackage[english]{babel}
\usepackage{amsmath,amssymb,amsbsy,amstext, amsthm, simplewick}
\usepackage{hyperref}
\usepackage{tikz}
\usetikzlibrary{decorations.markings}
\usepackage{xcolor}

\usepackage{caption}

\usepackage{amsfonts}
\usepackage{amssymb}
\usepackage{upgreek}
\usepackage{simplewick}
 \usepackage{exscale,relsize}
\usepackage{mathtools}
\usepackage{comment}

\usepackage[margin=1cm,labelfont={sf,bf,scriptsize},textfont={sf,scriptsize}]{caption}

\def\br{{\bf r}}

\def \la {\langle}
\def \ra {\rangle}

\def\nn{{\nonumber}}

\def\be#1\ee{\begin{align}#1\end{align}}

\begin{document}

\title{Black hole bulk-cone singularities
}

\author{Matthew Dodelson$^{a}$,}
\author{Cristoforo Iossa$^{b,c,d}$,}
\author{Robin Karlsson$^{a}$,}
\author{Alexandru Lupsasca$^{e}$, and}
\author{Alexander Zhiboedov$^a$}

\affiliation{$^a$CERN, Theoretical Physics Department, CH-1211 Geneva 23, Switzerland}
\affiliation{$^b$Section de Math\'{e}matiques, Universit\'{e} de Gen\`{e}ve, 1211 Gen\`{e}ve 4, Switzerland}
\affiliation{$^c$SISSA, via Bonomea 265, 34136 Trieste, Italy}
\affiliation{$^d$INFN, sezione di Trieste, via Valerio 2, 34127 Trieste, Italy}
\affiliation{$^e$Department of Physics \& Astronomy, Vanderbilt University, Nashville TN 37212, USA}

\abstract{Lorentzian correlators of local operators exhibit surprising singularities in theories with gravity duals.
These are associated with null geodesics in an emergent bulk geometry.
We analyze singularities of the thermal response function dual to propagation of waves on the AdS Schwarzschild black hole background.
We derive the analytic form of the leading singularity dual to a bulk geodesic that winds around the black hole.
Remarkably, it exhibits a boundary group velocity larger than the speed of light, whose dual is the angular velocity of null geodesics at the photon sphere.
The strength of the singularity is controlled by the classical Lyapunov exponent associated with the instability of nearly bound photon orbits.
In this sense, the bulk-cone singularity can be identified as the universal feature that encodes the ubiquitous black hole photon sphere in a dual holographic CFT.
To perform the computation analytically, we express the two-point correlator as an infinite sum over Regge poles, and then evaluate this sum using WKB methods.
We also compute the smeared correlator numerically, which in particular allows us to check and support our analytic predictions.
We comment on the resolution of black hole bulk-cone singularities by stringy and gravitational effects into black hole bulk-cone ``bumps''.
We conclude that these bumps are robust, and could serve as a target for simulations of black hole-like geometries in table-top experiments.
}

    \begin{flushleft}
 \hfill \parbox[c]{40mm}{CERN-TH-2023-192}
\end{flushleft}  
\maketitle

\section{Introduction}

Perturbative Lorentzian correlation functions of local operators in QFT develop singularities when operators can communicate by exchanging light-like signals \cite{Landau:1959fi,Coleman:1965xm}.\footnote{While these papers concern QFT in Minkowski space, the same statement is expected to be true in curved space as well, see e.g. \cite{Poisson:2003nc}.} A striking feature of holographic theories is that communication can proceed via an emergent bulk and in this way new singularities develop \cite{Hubeny:2006yu,Gary:2009ae,Maldacena:2015iua}, through which the bulk geometry can be reconstructed \cite{Engelhardt:2016wgb,Engelhardt:2016crc}. This emergent bulk communication channel cannot be faster than the boundary one \cite{Gao:2000ga}. In this way boundary causality is preserved.

The simplest singularity arises in the two-point function $\langle {\cal O}(x_1) {\cal O}(x_2) \rangle$ when the points become light-like separated, $(x_1-x_2)^2 =0$. A holographic analog of this light-cone singularity is the  \emph{bulk-cone singularity} \cite{Hubeny:2006yu}. It occurs when the boundary points are connected by a null geodesic in the bulk. In pure AdS, the two cones coincide, but they differ in non-trivial backgrounds. The existence of a nearly null geodesic in the bulk does not always lead to a singularity in the correlator \cite{Fidkowski:2003nf,Horowitz:2023ury,Festuccia:2005pi,Kolanowski:2023hvh},  but for the one-sided correlators we will consider these subtleties do not arise.\footnote{As opposed to the situation considered in the present paper, in the two-sided case the two points cannot exchange light-like signals.}

\begin{figure}[h]
    \centering
\includegraphics[width=0.75\textwidth]{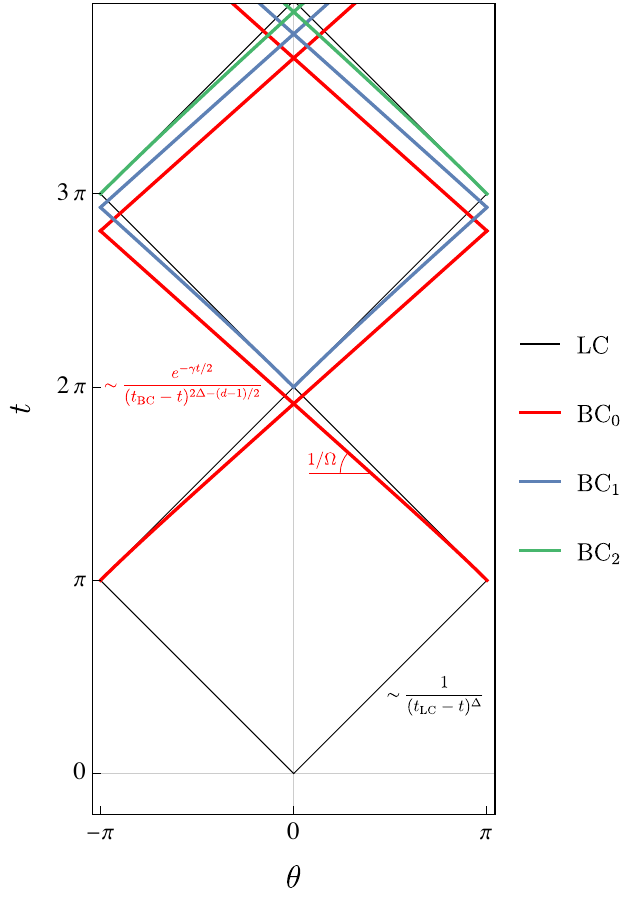}
    \caption{The structure of bulk-cone singularities in AdS Schwarzschild. The black line ${\rm LC}$ corresponds to the ordinary light-cone with the leading behavior controlled by the light-cone OPE ${1 \over (t_{{\rm LC}}-t)^\Delta}$. The  red line ${\rm BC}_0$ denotes a singularity due to a null geodesic in the bulk which wraps around the photon sphere, and its functional form is computed in this paper. It is more singular than the light-cone for $\Delta > {d-1 \over 2}$ and is given by ${1 \over (t_{{\rm BC}}-t)^{2\Delta-{d-1 \over 2}}}$. The effective group velocity of the bulk-cone singularity $\Omega>1$ is related to the angular velocity of null geodesics at the photon sphere. The strength of the bulk-cone singularity decays with time as $e^{- \gamma t/2}$, where $\gamma$ is the Lyapunov exponent of geodesics at the photon sphere.}
    \label{fig:bulkConeInt}
\end{figure}

\begin{figure}[h]
  \centering
  \begin{subfigure}{0.3\textwidth}
    \centering
    \includegraphics[width=\linewidth]{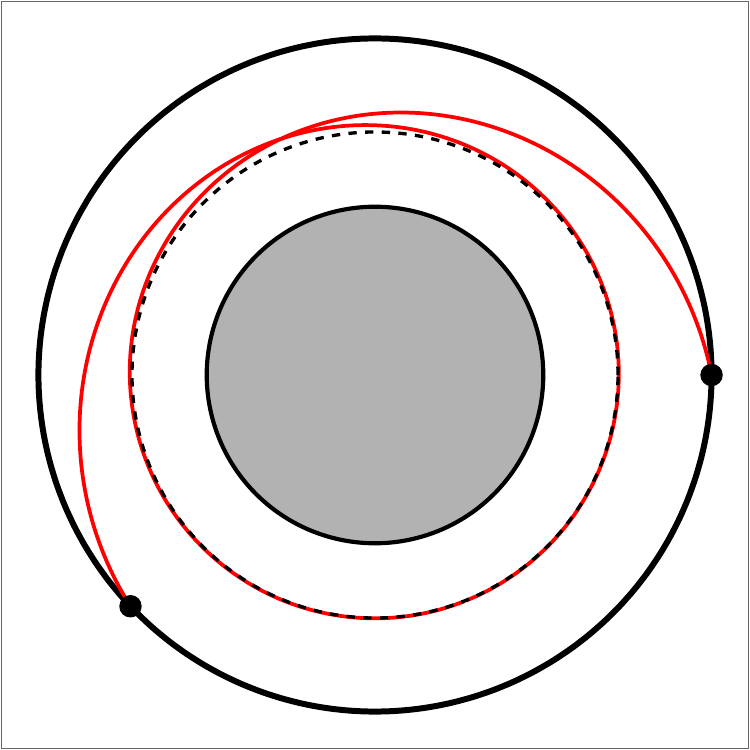}
    \caption{No bounce ${\rm BC}_0$}
        \label{fig:nobouncefig}
  \end{subfigure}
  \hfill
  \begin{subfigure}{0.3\textwidth}
    \centering
    \includegraphics[width=\linewidth]{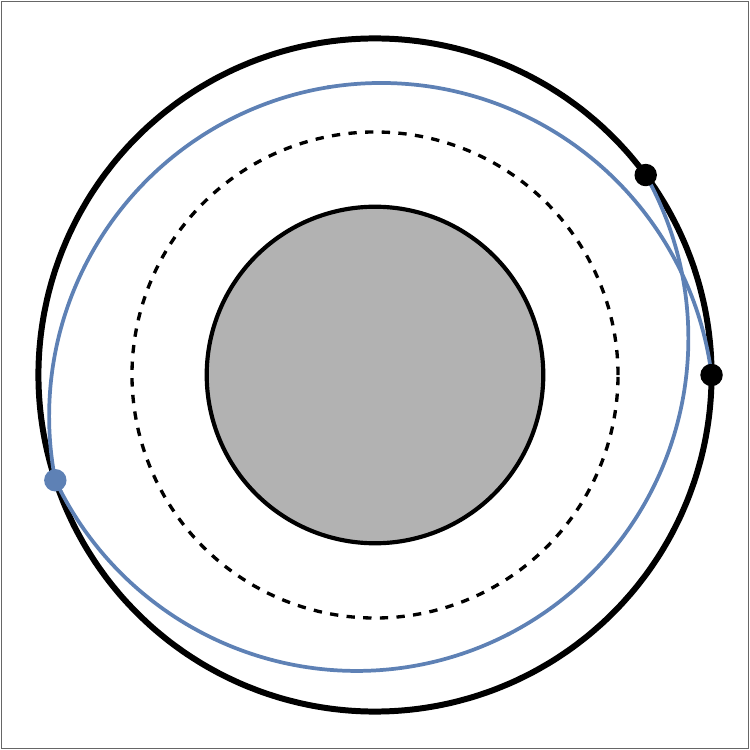}
    \caption{One bounce ${\rm BC}_1$}
\label{fig:1bouncefig}
  \end{subfigure}
  \hfill
  \begin{subfigure}{0.3\textwidth}
    \centering
    \includegraphics[width=\linewidth]{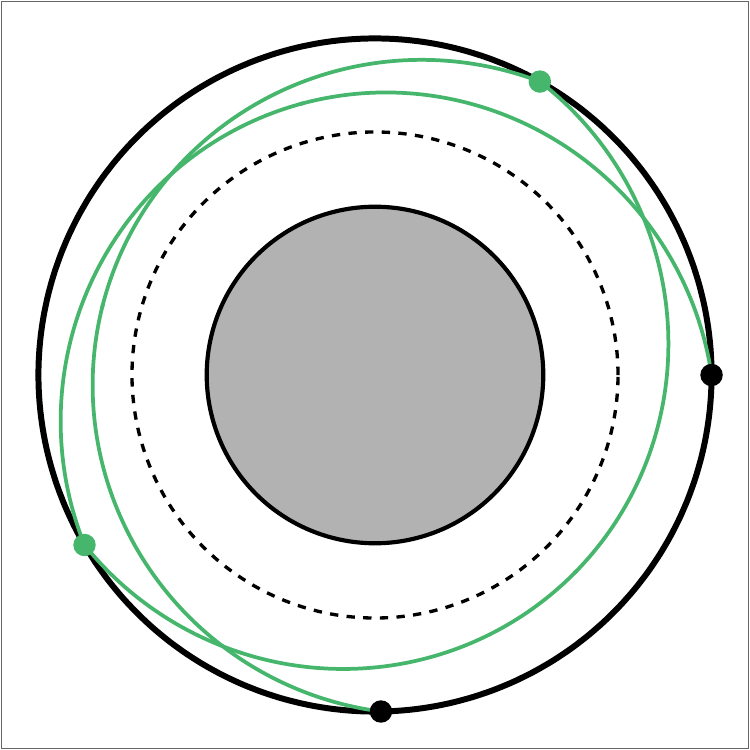}
    \caption{Two bounces ${\rm BC}_2$}
    \label{fig:2bouncefig}
  \end{subfigure}
  \caption{Null geodesics in AdS Schwarzschild ($d=4$, $\mu=1$). The gray shaded region corresponds to $r<r_s$, where $r_s$ is the Schwarzschild radius. The black dashed curve is the photon sphere. Black points denote the beginning and the end-point of the null geodesics that we are interested in. Colored points denote bounces. Given a bulk-cone singularity with no bounces at $(t,\theta)$, there will be an $(n-1)$-bounce bulk cone singularity at $(n t, n \theta)$.}
      \label{fig:bulkGeodesics}
\end{figure}

In this paper we derive the singularities of the thermal two-point function of a holographic CFT on $S^1 \times S^{d-1}$ in the black hole phase. These \emph{black hole bulk-cone singularities} were first conjectured in \cite{Hubeny:2006yu,Dodelson:2020lal} by analyzing the geodesic approximation, where the dimension of the probe operator becomes large. They have a remarkable feature that the emerging bulk-cone group velocity is larger than 1.\footnote{Similar effects of ``faster than light propagation'' have been observed experimentally in a variety of resonant media \cite{boyd2009controlling}.} It is controlled by the angular velocity $\Omega$ of null geodesics at the photon sphere,\footnote{To observe this effect it is important that we consider a CFT on $S^{d-1}$, since it is absent in infinite volume $\mathbb{R}^{d-1}$. On the other hand, we expect the effect to be present when a holographic CFT is put on other positively curved spatial manifolds as well.} see Figure \ref{fig:bulkConeInt} and Figure \ref{fig:bulkGeodesics}. As can be seen from Figure \ref{fig:bulkConeInt} this effect is perfectly consistent with boundary causality because it appears with some delay. We will also see that the strength of the singularity captures the Lyapunov exponent $\gamma$ of null geodesics in the vicinity of the photon sphere. In this sense, the bulk-cone singularity represents the encoding of the critical parameters $(\Omega,\gamma)$ of the photon sphere in the black hole hologram.

Surprisingly little is known about the structure of correlators close to the black hole bulk-cone singularities. This is the problem we address in the present paper. Our work can be viewed as the AdS analog of \cite{Dolan:2011fh} (see also \cite{Buss:2017vud,Casals:2016qyj,Zenginoglu:2012xe}), in which the structure of singularities was worked out for the two-point function of massless fields on the Schwarzschild background in four-dimensional asymptotically flat spacetime.

One important difference between the boundary-cone and bulk-cone singularities is that the former are universal and appear in every theory, whereas the latter are present in the strong coupling limit only.\footnote{It has been observed that instanton corrections can sometimes mimic strong coupling singularities \cite{Maldacena:2015iua}. It would be interesting to understand this better, especially in the thermal context (see \cite{Rey:2005cn} for a related discussion).} In particular, moving away from the strict 't Hooft limit and taking into account finite $\lambda$ and finite $N$ effects they are expected to become \emph{bulk-cone bumps} \cite{Maldacena:2015iua,Dodelson:2020lal}. As such they represent a simple, universal and robust feature of emergence of black hole-like geometries in holographic systems.

In this paper we analyze black hole bulk-cone singularities in the gravity approximation. In particular, we analytically compute the leading form of the singularity. 
The key observation is that the singularities are captured by a saddle point computation at complex spin. More precisely, we express the correlator as a sum over Regge poles and then use WKB methods to evaluate this sum to obtain the leading singularity.

Let us now present an overview of our final result. We consider the retarded two-point function $G_R(t, \theta) \equiv$  $i \theta(t) \langle [{\cal O}(t, \theta), {\cal O}(0,0)]\rangle_{S^1 \times S^{d-1}}$. The leading singular behavior associated  with the no-bounce null bulk geodesics, see Figure \ref{fig:nobouncefig}, is given by the formula
\be
&G_R(t, \theta) 
\propto \frac{(u(t)^2-1)^{\Delta -d/2}  }{(u(t))^{2 \Delta -{d-1 \over 2}} \sqrt{T'(u(t))} (\sin \theta)^{\frac{d-2}{2}}} \\
&\times\sum_{j=1}^\infty (-1)^{jd} {\rm Im} \left[ {1 \over \Big(t_{\text{BC}}(2 \pi j  + \theta) - t + i 0 \Big)^{2 \Delta -{d-1 \over 2}}} +  {e^{i \pi {d-2 \over 2}} \over \Big(t_{\text{BC}}(2 \pi j  - \theta) - t + i 0 \Big)^{2 \Delta -{d-1 \over 2}}} \right] \notag,
\ee
where $\pi > \theta >0$, and the omitted numerical pre-factor can be found in \eqref{eq:GWleadBC} and \eqref{eq:relRandW}. \\
\indent We now explain the basic elements of this formula.
\begin{itemize}
    \item $d$ is the dimensionality of the boundary CFT, and $\Delta$ is the scaling dimension of the scalar primary operator ${\cal O}$, related to the mass of the AdS dual bulk field $\phi$ as $\Delta = {d \over 2} + \sqrt{{d^2 \over 4} + (m R_{AdS})^2}$.
    \item The formula should be understood as a prediction for the leading singularities of the correlator close to $t=t_{\text{BC}}(2 \pi j \pm \theta)$ at fixed $\pi > \theta >0$. Away from the singularities there are subleading corrections which we do not compute in the present paper.
    \item Consider a null geodesic in the bulk that starts at the boundary point $(0,0)$ and ends at the boundary point $(T,\Theta)$. $t_{\text{BC}}(\Theta) \equiv T$ is the time it takes the bulk geodesic to traverse an angle $\Theta$.
     \item The $j$ sum is over winding geodesics that end at the same spatial point on the boundary as a result of the periodicity $\theta \sim \theta + 2 \pi$. The two terms in the brackets correspond to left and right-moving geodesics.
    \item $u(T) = {d \Theta \over d T} $ is the effective angular velocity of the bulk geodesic, with $1\le u\le \Omega$, where $\Omega$ is the angular velocity of circular null geodesics at the photon sphere, see Figure \ref{fig:u(t)}.
    \item At late times $T'(u(t)) =\left.\frac{d T}{d u}\right|_{u=u(t)} \sim e^{\gamma t}$ effectively measures the classical Lyapunov exponent $\gamma$ at the photon sphere, see Figure \ref{fig:LogT}. 
\end{itemize}  
We also computed the smeared correlator $G_R(t, \theta)$ numerically as in \cite{Buss:2017vud}, see Figure \ref{fig:GR3d} and Figure \ref{fig:GR4d}, which in particular allows us to check and support our analytic predictions. In addition, we observe singularities associated with bouncing geodesics, see Figure \ref{fig:1bouncefig} and Figure \ref{fig:2bouncefig}. 

The plan of the paper is as follows:
\begin{itemize}
    \item In Section \ref{sec:infinitevolume}, we study thermal correlators at infinite volume and reproduce the expected light-cone singularity from doubly infinite sums over Regge poles and Matsubara frequencies. This serves as a useful example before proceeding to the more interesting case of finite volume.
    \item In Section \ref{sec:eikonalcomplexspin}, we turn to finite volume thermal correlators and compute the eikonal spectrum of quasi-normal modes for both real and imaginary spin using WKB methods. We further comment on the flat space limit where, in particular, the spectrum at imaginary spin smoothly approaches the well-studied eikonal spectrum of black holes in asymptotically flat spacetime. 
    \item Section \ref{bulkcone} contains the main results of this paper. We begin by deriving the representation of the Euclidean thermal correlator at finite volume as a sum over Regge poles and Matsubara frequencies in any dimension. This can be analytically continued to real time in order to explore the singularity structure of Wightman correlators. Using the spectrum and residues found in Section \ref{sec:eikonalcomplexspin}, we obtain the light-cone singularity as well as the bulk-cone singularities present at finite volume. 
    \item Section \ref{sec:numerics} is devoted to numerical computations of retarded correlators in $d=3$ and $d=4$. We successfully match the location of the bulk cone singularities, their shape, and their relative strength with the analytical predictions from the previous sections. 
    \item In Section \ref{sec:string}, following \cite{Dodelson:2020lal,Amati:1987uf} we comment on stringy and gravitational corrections to our results. In particular, the singularities are smoothed out into finite width bumps. 
    \item In Section \ref{sec:prvsbh}, we review existing bounds on ultra-compact objects (different from black holes and possibly possessing a photon ring) in four-dimensional asymptotically flat spacetime.
\end{itemize}

We conclude and discuss open questions in Section \ref{sec:conclusions}. Appendix \ref{reggebounds} contains bounds on Regge poles from the wave equation. As a useful example of the methods described in this paper, the BTZ black hole is considered in Appendix \ref{app:BTZ}. In particular, both at  infinite and finite volume, the correlators written as sums over Regge poles and Matsubara frequencies are explicitly shown to reproduce known results. In Appendix \ref{app:causalitysphere} we derive causality constraints on the retarded two-point function on the sphere in momentum space. In Appendix \ref{app:latetimelightcone} we discuss the structure of the leading light-cone singularity of the two-point function at finite temperature in $d>2$. In Appendix \ref{sec:Signatures} we review observational signatures of astrophysical black holes.
\\

\noindent \emph{Note added:} 
Signatures of the photon sphere in the thermal two-point function in the eikonal limit $\omega, \ell \gg 1$, $\ell/\omega$ fixed were recently discussed in \cite{Hashimoto:2019jmw,Hashimoto:2023buz,Riojas:2023pew,riojas}. We consider this regime in Section \ref{sec:realspineikonal}.

\section{The light cone at infinite volume}
\label{sec:infinitevolume}
In this section, we derive a representation of the holographic thermal two-point correlator at infinite volume as a sum over Regge poles and Matsubara frequencies. We then use it to obtain the leading light-cone singularity of the Lorentzian thermal two-point function on $S^1_{\beta}\times \mathbb{R}^{d-1}$ in holographic theories. This section serves us a warm-up to the more interesting case of $S^1_{\beta}\times S^{d-1}$, which we consider in the next section.

\subsection{Holographic thermal two-point function}

We consider a scalar field in the background of a $(d+1)$-dimensional AdS black brane. The metric takes the form 
\begin{align}
ds^2=-f(r)\, dt^2+f(r)^{-1}\, dr^2+r^2\, d\vec{x}^2,\hspace{10 mm}f(r)=r^2-\frac{1}{r^{d-2}}.
\end{align}
We have chosen to work in units in which the horizon is at $r=1$ and the inverse temperature is $\beta=4\pi/d$. The AdS boundary is at $r=\infty$. The equation of motion $(\Box-m^2)\phi=0$ is conveniently written in terms of a new field $\psi$ defined by \cite{Festuccia:2008zx}
\begin{align}
\phi(t,\vec{x},r)=e^{-i\omega t+i\vec{k}\cdot \vec{x}}r^{-\frac{d-1}{2}}\psi_{\omega k}(r), ~~~~ k \equiv |\vec k|.
\end{align}
The radial part of the wave equation then becomes 
\begin{align}\label{waveeqbrane}
(-\partial_z^2+V(z)-\omega^2)\psi_{\omega k}(z)=0,
\end{align}
where we have introduced a new coordinate $z$ via
\be
dz=-\frac{dr}{f(r)},
\ee
such that $z=0$ corresponds to the AdS boundary and $z=\infty$ to the black hole horizon.
The potential $V(z)$ is given by
\begin{align}\label{potentialinf}
V(z)=f(r)\left(\frac{k^2}{r^2}+\nu^2-\frac{1}{4}+\frac{(d-1)^2}{4r^d}\right),
\end{align}c
and it is a monotonically increasing function of $r$ when $k$ is real. The conformal dimension of the boundary operator ${\cal O}(x)$ dual to the bulk field $\phi$ is 
\be
\Delta=\frac{d}{2}+\nu. 
\ee
\indent The retarded correlator is computed by specifying ingoing boundary conditions at the horizon \cite{Son:2002sd}, 
\begin{align}\label{eq:bdyInfVol}
\psi_{\omega k}(z)\sim e^{i\omega z},\hspace{10 mm}z\to \infty.
\end{align}
Near the boundary, the solution behaves as 
\begin{align}\label{boundarybehavior}
\psi_{\omega k}(z)\sim \mathcal{A}(\omega,k) z^{\frac{1}{2}-\nu}+\mathcal{B}(\omega,k)z^{\frac{1}{2}+\nu},\hspace{10 mm}z\to 0.
\end{align}
The retarded Green's function is then given by
\begin{align}\label{grdef}
G_R(\omega,k)=\frac{\mathcal{B}(\omega,k)}{\mathcal{A}(\omega,k)},
\end{align} 
where we have introduced
\be
\label{eq:retflat}
G_R(t,\vec x) &\equiv i \theta(t) \langle [{\cal O}(t, \vec x), {\cal O}(0,0)]\rangle_{S^1 \times \mathbb{R}^{d-1}}\\
&= {1 \over (2 \pi)^d}\int_{-\infty}^{\infty} d\omega\, d^{d-1} \vec k\,  e^{- i \omega t + i \vec k \cdot \vec x} G_R(\omega,|\vec k|) \ . 
\ee
We will be interested in the leading light-cone singularity of the thermal two-point function. This can be readily derived using WKB methods at large $\omega$ and $k$, see \cite{Festuccia:2005pi,Festuccia:2008zx,festuccia2007black}. However, for our purposes of understanding bulk-cone singularities at finite volume an alternative method via re-summation of Regge poles will turn out to be more useful. We present this method below.

\subsection{Regge poles and the Wightman function}\label{Sec:ReggeInfVol}

In fact, we will find it more convenient to compute the Wightman two-point function, and then use it to compute the retarded two-point function \eqref{eq:retflat} if necessary.

The real-time Wightman function in a finite temperature CFT on $S^1_{\beta}\times \mathbb{R}^{d-1}$ is defined as follows
\begin{align}
G_W(t,x)=\frac{1}{Z}\text{Tr}\left(e^{-\beta H}\mathcal{O}(t,\vec{x})\mathcal{O}(0)\right),\hspace{10 mm}x\equiv |\vec{x}|.
\end{align}
Recall that the Wightman function can be obtained from the Euclidean correlator by analytic continuation, 
\begin{align}
\label{eq:EtoW}
G_W(t,x)=\lim_{\epsilon\to0}G_E(\tau=\epsilon+it,x).
\end{align}
Moreover, the Euclidean correlator admits a Fourier decomposition (see e.g.\ \cite{Meyer:2011gj}),
\begin{align}\label{eq:GEInfVol}
G_E(\tau,x)={1 \over \beta} \sum_{n=-\infty}^{\infty}e^{i\zeta_n\tau}\int \frac{d^{d-1}\vec{k}}{(2\pi)^{d-1}}\, e^{i\vec{k}\cdot \vec{x}}G_R\left(i|\zeta_n|,k\right),\hspace{10 mm}k\equiv |\vec{k}|,
\end{align} 
where $G_R$ is the retarded Green's function. Here the Matsubara frequencies are given by
\begin{align}\label{matsubara}
\zeta_n=\frac{2\pi n}{\beta} \ ,
\end{align}
and $G_R(i\zeta_n,k)=G_E(\zeta_n,k)$. The expression above is manifestly KMS invariant, $G_E(\beta - \tau,x) = G_E(\tau, x)$. 

Notice that we cannot use \eqref{eq:GEInfVol} directly to compute the Lorentzian correlator. The reason is that plugging \eqref{eq:GEInfVol} into \eqref{eq:EtoW} leads to a divergent sum over Matsubara frequencies. To overcome this difficulty we use a trick familiar from Regge theory, and analytically continue the representation above in $k$. To achieve this we first note that the potential (\ref{potentialinf}) is an analytic function of $k^2$, so that the retarded two-point function computed from it is automatically analytic in $k$ and also obeys $G_R(\omega,k)=G_R(\omega,-k)$. It follows that we can extend the $k$ integral to the full real line,
\be
   \int \frac{d^{d-1}\vec{k}}{(2\pi)^{d-1}}e^{i\vec{k}\cdot \vec{x}}G_R(i|\zeta_n|,k) &= \frac{1}{(2 \pi)^{d-1 \over 2}x^{\frac{d-3}{2}}} \int_{0}^{\infty} dk\, k^{\frac{d-1}{2}} J_{\frac{d-3}{2}}(k x)G_R(i|\zeta_n|,k)\notag\\
   &=\frac{1}{2 (2 \pi)^{d-1 \over 2}x^{\frac{d-3}{2}}} \int_{-\infty}^{\infty} dk\, k^{\frac{d-1}{2}} H^{(1)}_{\frac{d-3}{2}}(k x)G_R(i|\zeta_n|,k),
\ee
where $J_\alpha(x)$ and $H_\alpha^{(1)}(x)$ are the Bessel and Hankel functions of first kind respectively.
When evaluating $k^{\frac{d-1}{2}} H^{(1)}_{\frac{d-3}{2}}(k x)$ for $k<0$ the correct continuation prescription is $k \to k+ i 0$.

At large $k$, the Hankel function behaves as $e^{ikx}$, so for $x>0$ the integrand exponentially decays as $\text{Im }k\to +\infty$. We can therefore close the contour in the upper half plane, picking up singularities along the way,
\begin{align}\label{closedcontour}
G_E(\tau,x)=\frac{i\pi \beta^{-1}}{(2\pi)^{\frac{d-1}{2}}x^{\frac{d-3}{2}}}\sum_{n=-\infty}^{\infty}\sum_{m} e^{i\zeta_n \tau}k_{mn}^{\frac{d-1}{2}}H_{\frac{d-3}{2}}^{(1)}(k_{mn}x)\underset{\hspace{2mm}k\to k_{mn}}{\text{Res}}G_R(i|\zeta_n|,k),
\end{align}
where the sum over $m$ runs over poles $k_{mn}$  of $G_R(i|\zeta_n|,k)$ in the upper half $k$-plane.
From (\ref{grdef}), we see that the poles of $G_R(\omega,k)$ are determined by the equation $\mathcal{A}(\omega,k)=0$. One familiar way to think about this equation is to fix $k$, and to look for solutions $\omega_m(k)$. These solutions are the frequencies of quasi-normal modes.  In the present context, we fix $\omega = i |\zeta_n|$ and look for solutions $k_m(\omega)$. We call these solutions {\it thermal Regge poles}. From (\ref{closedcontour}), the spectrum and residues of Regge poles are sufficient information to compute the Euclidean Green's function. 

As we will see shortly, the utility of \eqref{closedcontour} is that it can be directly used to analyze the singularities of the Lorentzian correlator via the analytic continuation $e^{i\zeta_n \tau} \to e^{-\zeta_n t} $. 

\subsection{Eikonal spectrum of thermal Regge poles}

\indent The singularities of the two-point function are controlled by the high energy and momentum asymptotics of the sum \eqref{closedcontour}. Therefore we need to understand the structure of thermal Regge poles in the eikonal limit where $|\zeta_n|\to \infty$ and $|k_m(i|\zeta_n|)|\to \infty$.

\begin{figure}
\centering
\begin{tikzpicture}[scale=1.4,domain=0.01:7.5,samples=500]

  \draw[->] (-.2,0) -- (7.5,0) node[right] {$z$}; g
  \draw[->] (0,-2.1) -- (0,2.1) node[above] {$V^{k=ip}_{\text{eik}}(z)$};

  \draw[red,very thick]    plot (\x,{1.9*(-6*\x/5-36*\x^3/125)/sinh(6*\x/5)});
    \draw [dashed] (7.5,-1.1) -- (0,-1.1) node[left,blue] {$-u^2$};
        \draw [dashed] (7.5,-1.8) -- (0,-1.9) node[left,blue] {$-1$};
      \draw [dashed] (3.17,-1.1) -- (3.17,0) node[above ,blue] {$z_T$};;
\end{tikzpicture}
 \caption{The eikonal potential (\ref{veikinf}) for imaginary $k=ip$ has one real turning point $z_T$. The minimum of the potential is at the boundary $z=0$. \label{potentialinfvol}}
\end{figure}
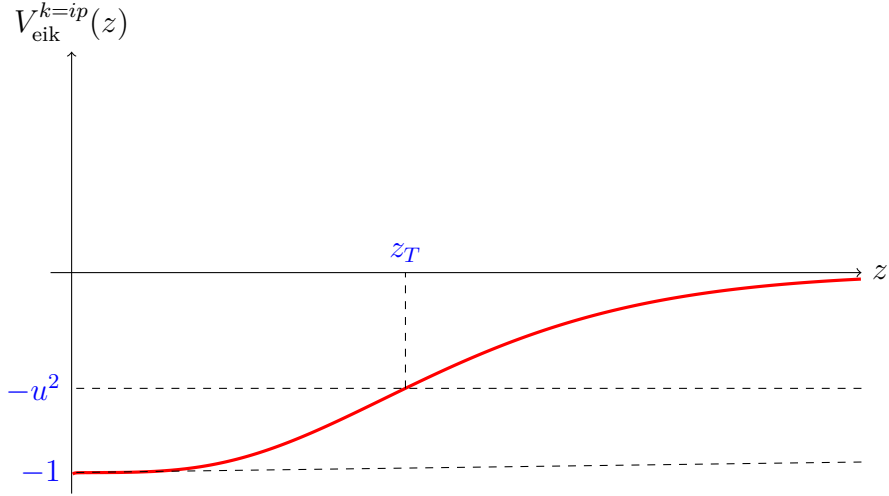

In Appendix \ref{reggebounds}, we show that for positive imaginary $\omega$, thermal Regge poles in the upper half plane must lie in the sector $\frac{\pi}{4}<\text{Arg}(k_m(\omega))<\frac{3\pi}{4}$. In fact, we found numerically that all the Regge poles for positive imaginary $\omega$ are at imaginary $k$, see Figure \ref{fig:reggepoles}. Therefore, let us define $k=ip$ and $\omega=ipu$ and then take $p\to \infty$. Then away from the boundary $z\gg 1/p$, the wave equation (\ref{waveeqbrane}) reduces to 
\begin{align}
(\partial_z^2+p^2\kappa^2(z))\psi(z)=0,\hspace{5 mm}\kappa(z)=\sqrt{-V^{k=ip}_{\text{eik}}(z)-u^2},
\end{align}
where we have defined the eikonal potential 
\begin{align}\label{veikinf}
V^{k=ip}_{\text{eik}}(z)=\frac{1}{r^d}-1.
\end{align}
This potential admits bound states which appear as Regge poles on the imaginary $k$ axis, and we wish to compute the locations of these poles and their residues. \\
\indent We take $0<u<1$, so that there is a single turning point located at $r_T=(1-u^2)^{-1/d}$, as depicted in Figure \ref{potentialinfvol}. This turning point approaches the boundary as $u\to1$ and the horizon as $u\to0$. In the WKB approximation, the ingoing solution takes the following form near the horizon, 
\begin{align}
\psi(z)\sim \frac{1}{\sqrt{|\kappa(z)|}}e^{-p\int_{z_T}^{z}\, dz' |\kappa(z')|},\hspace{10 mm}z>z_T.
\end{align}
Note that this is exponentially decaying at infinity, as is appropriate for a bound state. Applying the WKB connection formula then gives the solution in the classically allowed region, 
\begin{align}\label{smallzinf}
\psi(z)\sim \frac{e^{\frac{i\pi}{4}-iS(0,z_T)}}{\sqrt{\kappa(z)}}e^{ip\int_{0}^{z}\, dz' \kappa(z')}+\frac{e^{-\frac{i\pi}{4}+iS(0,z_T)}}{\sqrt{\kappa(z)}}e^{-ip\int_{0}^{z}\, dz' \kappa(z')},\hspace{10 mm}z<z_T.
\end{align}
Here we have defined the WKB action 
\begin{align}
S(z_a,z_b)=p\int_{z_a}^{z_b}dz'\, |\kappa(z')|.
\end{align}

\indent The WKB solution (\ref{smallzinf}) does not yield the correct boundary asymptotics (\ref{boundarybehavior}). The reason is that the eikonal potential is no longer dominant at $z$ of order $1/p$. In this region the potential (\ref{potentialinf}) can be approximated as
\begin{align}
V(z)\sim \frac{\nu^2-\frac{1}{4}}{z^2}-p^2, \hspace{10 mm}z\sim \frac{1}{p}\ll 1.
\end{align}
The solutions are Hankel functions, whose coefficients can be fixed by matching to the WKB solution (\ref{smallzinf}) at $z\gg 1/p$. We find 
\begin{align}
\psi(z)&\sim i\sqrt{\frac{\pi  p z}{2}}e^{\frac{i\pi \nu}{2}-iS(0,z_T)}H^{(1)}_{\nu}(p\sqrt{1-u^2}z)\\
&\hspace{5 mm}-i\sqrt{\frac{\pi  p z}{2}} e^{-\frac{i\pi \nu}{2}+iS(0,z_T)}H^{(2)}_{\nu}(p\sqrt{1-u^2}z),\hspace{5 mm}z\sim \frac{1}{p}.\notag
\end{align}
The retarded Green's function (\ref{grdef}) can then be read off from the asymptotics (\ref{boundarybehavior}),
\begin{align}\label{eq:GRInf}
G_R(\omega,k)=\frac{\Gamma(-\nu)}{\Gamma(\nu)}\left(\frac{\omega^2-k^2}{4}\right)^{\nu}\frac{\cos\left(S(0,z_T)+\frac{\pi\nu}{2}\right)}{\cos\left(S(0,z_T)-\frac{\pi\nu}{2}\right)} .
\end{align}
In deriving this formula we assumed that $S(0,z_T)\gg 1$. Corrections to this formula are suppressed by $\frac{1}{S(0,z_T)}$.

\indent Now let us compute the spectrum of thermal Regge poles from (\ref{eq:GRInf}). The pole condition reduces to 
\begin{align}\label{polecondinf}
S(0,z_T)=\frac{\pi(\nu+1+2m)}{2},\hspace{10 mm}m=0,1,\ldots,
\end{align}
where $S(0,z_T)$ is given by 
\be 
S(0,z_T) &= \notag p\sqrt{1-u^2}\int_{r_T}^\infty dr\, \frac{\sqrt{1-\left(\frac{r_T}{r}\right)^d}}{r^2-r^{2-d}}\\
 &= \frac{\sqrt{\pi } p (1-u^2)^{\frac{1}{d}+\frac{1}{2}}\Gamma \left(1+\frac{1}{d}\right) \,
   _2F_1\left(1,\frac{1}{d};\frac{3}{2}+\frac{1}{d};1-u^2\right)}{2 \Gamma \left(\frac{3}{2}+\frac{1}{d}\right)}.
\ee 
As mentioned above, this calculation can be trusted for $m\gg1$. The residues of $G_R$ in \eqref{eq:GRInf} are found to be
\be\label{eq:resInf}
    \underset{\hspace{2mm}k\to k_m}{\text{Res}}G_R(\omega,k)=-\frac{\pi}{\nu\Gamma(\nu)^2}\left(\frac{\omega^2-k_m(\omega)^2}{4}\right)^{\nu}\left(\frac{\partial S(0,z_T)}{\partial k}\Big|_{k_m}\right)^{-1}.
\ee
In particular, as the turning point approaches the boundary, we have $u\to1$, so that we can expand the action as 
\begin{align}
S(0,z_T)= \frac{p\sqrt{\pi}2^{-\frac{1}{2}+\frac{1}{d}}\Gamma\left(1+\frac{1}{d}\right)}{\Gamma\left(\frac{3}{2}+\frac{1}{d}\right)}(1-u)^{\frac{1}{2}+\frac{1}{d}} +\ldots ,\hspace{10 mm}u\to 1.
\end{align}
Plugging into (\ref{polecondinf}), we can read off the low-lying spectrum
\be \label{eikpolesinf}
k_m(\omega)=\omega+\frac{i}{(-i\omega)^{\frac{d-2}{d+2}} }\left(\frac{\sqrt{\pi}\Gamma
   \left(\frac{3}{2}+\frac{1}{d}\right) m}{2^{\frac{1}{d}-\frac{1}{2}} \Gamma
   \left(1+\frac{1}{d}\right)}\right)^{\frac{2 d}{d+2}}+\ldots,
\ee
for $k_m\sim\omega$ and $|\omega|\gg m\gg 1$.  \\
\indent As mentioned above, thermal Regge poles and quasi-normal modes are equivalent descriptions of the poles of $G_R(\omega,k)$. We can invert the relation (\ref{eikpolesinf}) to find the spectrum of quasi-normal modes,
\be\label{eq:largepInf}
\omega_m(k)=k-i\frac{1}{(-ik)^{\frac{d-2}{d+2}}} \left(\frac{\sqrt{\pi}\Gamma \left(\frac{3}{2}+\frac{1}{d}\right) m}{2^{\frac{1}{d}-\frac{1}{2}} \Gamma
   \left(1+\frac{1}{d}\right)}\right)^{\frac{2 d}{d+2}}+\ldots,
\ee 
for $\omega_m\sim k$ and $|k|\gg m\gg 1$. 

Let us make several comments on this formula. First, note that $\text{Im }\omega_m(k)<\text{Im }k$, as required by causality at infinite volume \cite{Heller:2022ejw}. Second, the spectrum (\ref{eq:largepInf}) precisely agrees with the naive rotation of the real $k$ spectrum \cite{Festuccia:2008zx} to imaginary $k$.\footnote{Here we have corrected a minor numerical typo in \cite{Festuccia:2008zx}.} At finite volume, we will see that both of these properties are violated: there are modes with $\text{Im }\omega_m(\ell)>\text{Im }\ell$, where $\ell$ is the spin, and the imaginary spin modes are not obtained by simply rotating the real spin modes.

\subsection{The light-cone singularity}
Next we will see how the light-cone singularity is reproduced from the Regge expansion (\ref{closedcontour}). To this end, let us evaluate \eqref{closedcontour} on the Regge poles at the Matsubara frequencies, with the residues \eqref{eq:resInf}. Since we are interested in the singularities in position space, we approximate the sums by integrals for $\zeta_n\sim p_m(i|\zeta_n|)\gg1$,
\be\label{sumstointegrals}
{1 \over \beta}\sum_n \sum_m \simeq {1 \over 2 \pi}  \int_{0}^{\infty} d p\, {d m \over d p}\int_{-p}^{p} d \zeta = {1 \over 2 \pi} \int_{0}^{\infty} d p \int_{-p}^{p} d \zeta \,{i \over \pi} {d S(0, z_T) \over d (i p)} .
\ee
The last factor cancels the corresponding derivative in \eqref{eq:resInf}. We also use the large $p$ asymptotic of the Hankel function,
\begin{align}
H_{\frac{d-3}{2}}^{(1)}(i p x) \to \sqrt{\frac{2i}{\pi px}}e^{-\frac{1}{4}i \pi  d}e^{-p x}.   
\end{align}

Combining all the factors together, we get the following expression for the Euclidean two-point function,
\be 
G_E(\tau,x)=\frac{1}{ 2^{2\nu+1}\pi (2\pi x)^{\frac{d}{2}-1}\nu\Gamma(\nu)^2}\int_0^\infty dp \,\int_{-p}^{p}d\zeta\,e^{i\zeta \tau-p x}p^{\frac{d}{2}-1}(p^2-\zeta^2)^\nu .
\ee
Computing the leading light-cone singularity as $t \to x$, we get
\be
G_W(t,x)= {\Gamma(\Delta) \over \pi^{{d\over 2}} \Gamma(\Delta - {d \over 2})} {1 \over (x^2 - t^2)^{\Delta}} + ... \ ,
\ee
which correctly reproduces the expected leading light-cone singularity of the two-point function and serves as a consistency check for our computation. The factor ${\Gamma(\Delta) \over \pi^{{d\over 2}} \Gamma(\Delta - {d \over 2})}$ in front is simply an overall normalization of the operators. Notice that in contrast to $d=2$, see Appendix \ref{app:BTZ}, the leading light-cone singularity is the same as in the vacuum.

Equivalently, we could have used WKB methods to compute the large $\omega, k$ asymptotic for real $\omega$ and $k$ directly, from which the leading light-cone singularity trivially follows. The result above predicts that
\be
G_W(\omega, k) \simeq {2 \pi \left(\frac{\omega^2-k^2}{4}\right)^{\Delta - d/2} \over \Gamma(\Delta - {d \over 2}) \Gamma(\Delta + 1- {d \over 2})} ,
\ee
which indeed coincides with the results in \cite{Festuccia:2005pi,Festuccia:2008zx,festuccia2007black}, up to a factor $2 \Delta - d$ which is due to the difference in our definition of the retarded two-point function \eqref{grdef}.

\section{Eikonal spectrum of quasi-normal modes for complex spin}
\label{sec:eikonalcomplexspin}

Now let us turn to finite volume. In the next section we will compute the singularity structure of the correlator using the same method as at infinite volume. To do this we first need to understand the poles and residues of the retarded Green's function in the eikonal limit $\omega,\ell\to \infty$ with $\omega/\ell$ fixed. In this section we will focus on the quasi-normal modes rather than the Regge poles in order to make contact with prior work. The translation to Regge poles is straightforward, and will be done in Section \ref{bulkcone} when discussing the singularities. For complex spin, the answer is highly sensitive to the argument of $\ell$, and we will treat the cases of real $\ell$ and imaginary $\ell$ separately.\\
\indent We consider scalar field propagation in the AdS$_{d+1}$ Schwarzschild black hole with metric
\begin{align}
ds^2=-f(r)\, dt^2+f(r)^{-1}\, dr^2+r^2\, d\Omega_{d-1}^2, \hspace{10 mm}f(r)=r^2+1-\frac{\mu}{r^{d-2}}.
\end{align}
The black hole geometry dominates the canonical ensemble above the Hawking-Page transition, $\mu>2$ \cite{Witten:1998zw}.
The Schwarzschild radius is at 
\begin{align}
r_s^2+1-\frac{\mu}{r_s^{d-2}}=0.
\end{align}
The equation of motion simplifies after Fourier decomposition,
\be\label{fourierspin}
    \phi(t,\Omega_{d-1},r)= e^{-i\omega t}Y_{\ell\vec{m}}(\Omega_{d-1})r^{-\frac{d-1}{2}}\psi_{\omega\ell}(r),
\ee
where $Y_{\ell\vec{m}}$ are spherical harmonics on $S^{d-1}$. The wave equation takes the same form (\ref{waveeqbrane}) as at infinite volume, with the potential
\begin{align}\label{potential}
V(z)=f(r)\left(\frac{(\ell+\alpha)^2-\frac{1}{4}}{r^2}+\nu^2-\frac{1}{4}+\frac{(d-1)^2\mu}{4r^d}\right),\hspace{5 mm}\alpha \equiv \frac{d-2}{2}.
\end{align}

\subsection{Real spin and long-lived quasiparticles}
\label{sec:realspineikonal}

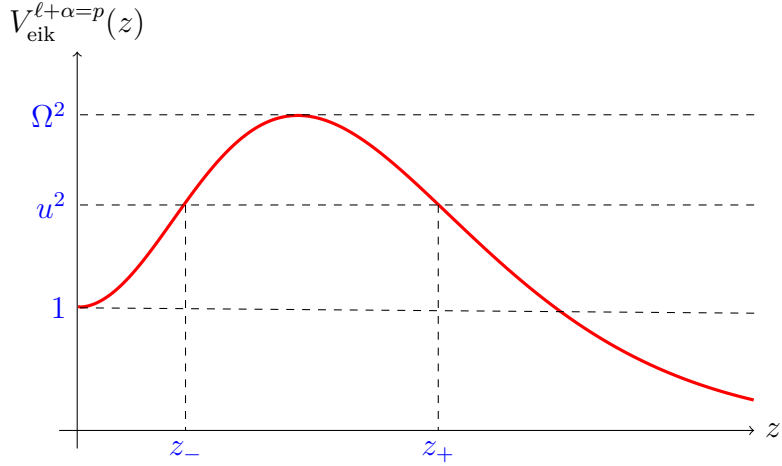
\begin{figure}
\centering
\begin{tikzpicture}[scale=1.2,domain=0.01:7.5,samples=500]

  \draw[->] (-0.2,0) -- (7.5,0) node[right] {$z$};
  \draw[->] (0,-.2) -- (0,4.2) node[above] {$V^{\ell+\alpha=p}_{\text{eik}}(z)$};

  \draw[red,very thick]    plot (\x,{1.5*\x*(1+\x^2)/sinh(1.1*\x)});
  \draw [dashed] (7.5,1.3) -- (0,1.36) node[left,blue] {$1$};
    \draw [dashed] (7.5,3.5) -- (0,3.5) node[left,blue] {$\Omega^2$};
      \draw [dashed] (7.5,2.5) -- (0,2.5) node[left,blue] {$u^2$};
    \draw [dashed] (1.2,2.5) -- (1.2,0) node[below,blue] {$z_-$};
      \draw [dashed] (4,2.5) -- (4,0) node[below,blue] {$z_+$};
\end{tikzpicture}
 \caption{The eikonal potential (\ref{veikrealspin}) for real spin has two real turning points $z_+$ and $z_-$ when $1<u<\Omega$. The height of the barrier is $\Omega^2$.\label{potentialreall}}
\end{figure}

Let us first review the case when the spin $\ell$ is large and real. The eikonal limit is defined by taking $p=\ell+\alpha$ and $\omega=pu$, with $p\to \infty$. Then for $z\gg 1/p$, the wave equation becomes
\begin{align}
(\partial_z^2+p^2\kappa^2(z))\psi(z)=0,\hspace{10 mm}\kappa(z)=\sqrt{u^2-V^{\ell+\alpha=p}_{\text{eik}}(z)},
\end{align}
where the eikonal potential is 
\begin{align}\label{veikrealspin}
V^{\ell+\alpha=p}_{\text{eik}}(z)=1+\frac{1}{r^2}-\frac{\mu}{r^{d}}.
\end{align}
As shown in Figure \ref{potentialreall}, there is a potential barrier separating the boundary from the horizon. The maximum of the potential is at the photon sphere 
\begin{align}
r_{\text{phot}}=\left(\frac{d\mu}{2}\right)^{\frac{1}{d-2}}.
\end{align}
The height of the barrier is 
\begin{align}
\Omega^2=1+\left(1-\frac{2}{d}\right)\left(\frac{2}{d\mu}\right)^{\frac{2}{d-2}} > 1,
\end{align}
which coincides with the velocity squared of null geodesics at the photon sphere. \\
\indent We work in the regime $1<u<\Omega$, so that there are two real turning points $z_{\pm}$ with $\kappa(z_{\pm})=0$ in the WKB region $z\gg 1/p$, as shown in Figure \ref{potentialreall}. The ingoing solution at the horizon is
\begin{align}
\psi(z)\sim \frac{1}{\sqrt{\kappa(z)}}e^{ip\int_{z_+}^{z}\, dz' \kappa(z')},\hspace{10 mm}z>z_+.
\end{align}
Assuming that $z_-$ and $z_+$ are sufficiently far apart, the behavior of $\psi(z)$ for $z<z_-$ can be computed by applying the standard WKB connection formulae twice\footnote{When the two turning points approach each other, this procedure is no longer valid and one should use a uniform approximation instead \cite{PhysRev.91.174,Berry:1972na}.}. One finds 
\begin{align}\label{smallzreall}
\psi(z)&\sim \frac{e^{S(z_-,z_+)}+\frac{1}{4}e^{-S(z_-,z_+)} }{\sqrt{\kappa(z)}}e^{ip\int_{z_-}^{z}dz'\kappa(z')}\notag\\
&\hspace{5 mm}-i\frac{e^{S(z_-,z_+)}-\frac{1}{4}e^{-S(z_-,z_+)}}{\sqrt{\kappa(z)}}e^{-ip\int_{z_-}^{z}dz'\kappa(z')},\hspace{ 5mm} z<z_-.
\end{align}
\indent The WKB solution (\ref{smallzreall}) can now be matched to Hankel functions in the same manner as at infinite volume,
\begin{align}
\psi(z)&\sim \sqrt{\frac{\pi i p z}{2}}\left(e^{S(z_-,z_+)}+\frac{1}{4}e^{-S(z_-,z_+)}\right) e^{\frac{i\pi \nu}{2}-iS(0,z_-)}H^{(1)}_{\nu}(p\sqrt{u^2-1}z)\\
&\hspace{5 mm}-\sqrt{\frac{\pi i p z}{2}}\left(e^{S(z_-,z_+)}-\frac{1}{4}e^{-S(z_-,z_+)}\right) e^{-\frac{i\pi\nu}{2}+iS(0,z_-)}H^{(2)}_{\nu}(p\sqrt{u^2-1}z),\hspace{5 mm}z\sim \frac{1}{p}.\notag
\end{align}
The retarded Green's function (\ref{grdef}) can then be read off from the asymptotics (\ref{boundarybehavior}), 
\begin{align}\label{grreall}
G_R(\omega,\ell)= \frac{\Gamma(-\nu)}{\Gamma(\nu)}\left(\frac{\omega^2-p^2}{4}\right)^{\nu}\frac{\cos\left(S(0,z_-)+\frac{\pi\nu}{2}\right)-\frac{i}{4}e^{-2S(z_-,z_+)}\sin\left(S(0,z_-)+\frac{\pi\nu}{2}\right)}{\cos\left(S(0,z_-)-\frac{\pi\nu}{2}\right)-\frac{i}{4}e^{-2S(z_-,z_+)}\sin\left(S(0,z_-)-\frac{\pi\nu}{2}\right)},
\end{align}
where in deriving this formula we have assumed that $S(z_-,z_+) \gg 1$ so that the turning points are not too close together. There are also corrections suppressed by $1/p$.
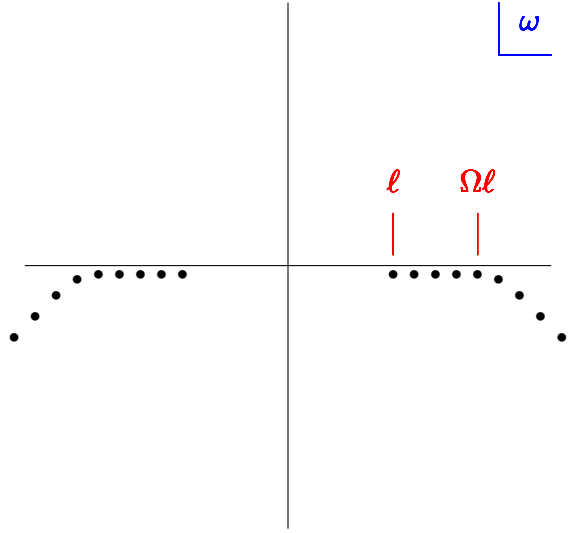
\begin{figure}
\centering
\begin{tikzpicture}[scale=1.4]
\draw (-2.5,0) -- (2.5,0);
\draw (0,-2.5) -- (0,2.5);
\foreach \Point in {(1,-.1),(1.2,-.1),(1.4,-.1),(1.6,-.1),(1.8,-.1),(2,-.15),(2.2,-.3),(2.4,-.5),(2.6,-.7),(-1,-.1),(-1.2,-.1),(-1.4,-.1),(-1.6,-.1),(-1.8,-.1),(-2,-.15),(-2.2,-.3),(-2.4,-.5),(-2.6,-.7)}{
    \node at \Point {\textbullet};
        \draw[red] (1,.1) -- (1,.5);
        \draw[red] (1.8,.1) -- (1.8,.5);
       \node[red] at (1.8,.8)  {$\Omega \ell$};
              \node[red] at (1,.8)  {$\ell$};
                   \node[blue] at (2.3,2.3)  {$\omega$};
                      \draw[blue,] (2.5,2) -- (2,2);
      \draw[blue] (2,2) -- (2,2.5);
       }
\end{tikzpicture}
 \caption{The spectrum of quasi-normal modes for real $\ell$. For $\ell <\text{Re }\omega<\Omega \ell$ the modes are quasiparticles with exponentially small imaginary part. For $\text{Re }\omega>\Omega \ell$ the modes go off into the complex plane at a finite angle.}   \label{realellqnms}
\end{figure}

\indent Using (\ref{grreall}), we can now compute the spectrum and residues. Recall that we are working in the regime where the tunneling action $S(z_-,z_+)$ is large. The pole condition therefore reduces to 
\begin{align}\label{polecondition}
S_m(0,z_-)=\frac{\pi}{2}(\nu+1+2m)-\frac{i}{4}e^{-2S_m (z_-,z_+)}+\ldots,\hspace{10 mm}m=0,1,\ldots,
\end{align}
which signifies a large number of quasiparticles with exponentially small decay rate \cite{Festuccia:2008zx,festucciathesis,Dodelson:2022eiz,Gannot:2012pb,Berenstein:2020vlp}. Note that we can only trust this formula for $m\gg 1$ in light of the comments above. 
This structure of QNMs as a function of an overtone number for fixed large spin $\ell$ has recently been discussed in \cite{Hashimoto:2019jmw,Hashimoto:2023buz,Riojas:2023pew}. These quasiparticles extend to $\omega=\Omega \ell$, and smoothly join onto a line of quasi-normal modes with order one imaginary part, as shown in Figure \ref{realellqnms}. The residues at the quasiparticles are 
\begin{align}\label{resreall}
 \underset{\hspace{2mm}\omega\to \omega_m}{\text{Res}}G_R(\omega,\ell)=-\frac{\pi}{\nu\Gamma(\nu)^2}\left(\frac{\omega_m^2-p^2}{4}\right)^{\nu}\frac{1}{T_m(0,z_-)},
\end{align}
where $ T_m(0,z_-)$ is the time traversed by a null geodesic from $z=0$ to $z=z_-$,
\begin{align}
T_m(0,z_-)&=\frac{dS(0,z_-)}{d\omega}\big|_{\omega=\omega_m}\notag\\
&=\int_{r(z_-)}^{\infty} \frac{dr}{f(r)}\, \frac{1}{\sqrt{1-\frac{p^2f(r)}{\omega_m^2 r^2}}}.
\end{align}\\
\indent The spectrum and residues simplify in the limit $m\ll p$, when the wavefunction is localized near the boundary of AdS. The action $S(0,z_-)$ can be evaluated explicitly in this limit,
\begin{align}
S(0,z_-)=\frac{\pi p}{2}(u-1),\hspace{10 mm}u\sim 1.
\end{align}
The quasi-normal mode spectrum (\ref{polecondition}) is then (recall that $p=\ell+\alpha$)
\begin{align}\label{omegan0}
\text{Re }\omega_m(\ell)=\ell+\Delta+2m,\hspace{10 mm} 1 \ll m\ll \ell,
\end{align}
which matches the spectrum of descendant operators in pure AdS. A more careful analysis, see e.g.\ \cite{Dodelson:2022eiz}, implies that the formula is correct starting from $m=0$. The residues (\ref{resreall}) become
\begin{align}\label{res0}
\underset{\hspace{2mm}\omega\to \omega_m}{\text{Res}}G_R(\omega,\ell)=-\frac{2}{\nu\Gamma(\nu)^2}(\ell m)^{\nu}.
\end{align}

We can compute the leading light-cone singularity from (\ref{res0}) using the results of \cite{Dodelson:2022eiz}. In particular see Formula (4.32) in that paper, noticing that the relationship between $P_J^{(d)}(\cos \theta)$ used there and Gegenbauer polynomials used here leads to $\underset{\hspace{2mm}\omega\to \omega_m}{\text{Res}}G_R(\omega,\ell) {\ell^{{d \over 2}-1} \over \Gamma(d/2)}=-c_{\text{norm}} c_{m,\ell}$. We conclude that the operators are normalized with $c_{\text{norm}} ={2 \Gamma(\Delta) \over \Gamma(d/2) \Gamma(\Delta-d/2)}$, so that the identity operator contributes as 
\begin{align}\label{boundarylc}
G_W(t,\theta) =  {2 \Gamma(\Delta) \over \Gamma(d/2) \Gamma(\Delta-d/2)}{1 \over 2^\Delta (\cos t - \cos \theta)^\Delta} + \dots \ . 
\end{align}
We will reproduce the same result in the next section by doing the computation at complex spin.

\subsection{Imaginary spin and the photon sphere}
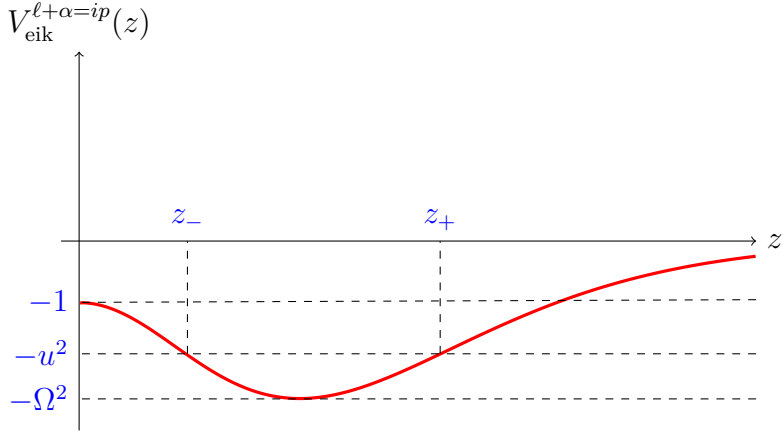
\begin{figure}
\centering
\begin{tikzpicture}[scale=1.2,domain=0.01:7.5,samples=500]

  \draw[->] (-0.2,0) -- (7.5,0) node[right] {$z$};
  \draw[->] (0,-2.1) -- (0,2.1) node[above] {$V^{\ell+\alpha=ip}_{\text{eik}}(z)$};

  \draw[red,very thick]    plot (\x,{-1.5*\x*(1+\x^2)/sinh(1.1*\x)/2});
  \draw [dashed] (7.5,-1.3/2) -- (0,-1.36/2) node[left,blue] {$-1$};
    \draw [dashed] (7.5,-3.5/2) -- (0,-3.5/2) node[left,blue] {$-\Omega^2$};
      \draw [dashed] (7.5,-2.5/2) -- (0,-2.5/2) node[left,blue] {$-u^2$};
    \draw [dashed] (1.2,-2.5/2) -- (1.2,0) node[above,blue] {$z_-$};
      \draw [dashed] (4,-2.5/2) -- (4,0) node[above,blue] {$z_+$};
\end{tikzpicture}
 \caption{The potential (\ref{veikimspin}) for imaginary spin has two real turning points when $1<u<\Omega$. The depth of the well is $\Omega^2$. \label{potentialiml}}
\end{figure}

\indent We now turn to the case of large imaginary spin, in which we define $\ell+\alpha=ip$ and $\omega=ipu$ and take $p\to\infty$. The wave equation (\ref{waveeqbrane}) becomes 
\begin{align}
(\partial_z^2+p^2\kappa^2(z))\psi(z)=0,\hspace{10 mm}\kappa(z)=\sqrt{-V^{\ell+\alpha=ip}_{\text{eik}}(z)-u^2},
\end{align}
where the eikonal potential is inverted in comparison to the real spin potential (\ref{veikrealspin}),
\begin{align}\label{veikimspin}
V_{\text{eik}}^{\ell+\alpha=ip}(z)=-1-\frac{1}{r^2}+\frac{\mu}{r^d}
\end{align}
The rotation from real to imaginary spin has converted the potential barrier into a potential well, as depicted in Figure \ref{potentialiml}. This potential admits bound states which appear as quasi-normal modes on the positive imaginary $\omega$ axis, and we wish to compute the locations of these poles and their residues. To this end, we will repeat the analysis of the previous subsection, pointing out several important differences along the way.\\
\indent Let us first discuss the regime $0<u<1$, where the eikonal potential (\ref{veikimspin}) has one real turning point $z_T$. In this case the analysis is identical to the infinite volume computation. The quasi-normal modes and residues are given by 
\begin{align}\label{finitevolumeordinary}
S(0,z_T)&=\frac{\pi(\nu+1+2m)}{2},\hspace{ 10 mm}m=0,1,\ldots,\\
\underset{\hspace{2mm}\omega\to \omega_m}{\text{Res}}G_R(\omega,\ell)&=-\frac{\pi}{\nu\Gamma(\nu)^2}\left(\frac{\omega_m(\ell)^2-(\ell+\alpha)^2}{4}\right)^{\nu}\left(\frac{\partial S(0,z_T)}{\partial \omega}\Big|_{\omega_m}\right)^{-1}.\label{finitevolumeordinaryres}
\end{align}
This formula captures the leading large $\ell$ behavior of the residues, with further corrections suppressed by $\frac{1}{\ell}$. It is valid when $S(0,z_T) \sim m \gg 1$.
\\\indent The more interesting regime is when $u$ is real with $1<u<\Omega$, so that there are two turning points at $z_{\pm}$. The wave equation is then real, and the ingoing boundary condition translates to normalizability of the solution at the horizon. The normalizable solution is
\begin{align}
\psi(z)=\frac{1}{\sqrt{|\kappa(z)|}}e^{-p\int_{z_+}^{z}dz'\,|\kappa(z')|},\hspace{5 mm}z>z_+.
\end{align}
\indent Next we would like to solve the connection problem from the region $z>z_+$ to the region $z<z_-$. In contrast to the case of real spin, we will not assume that the two turning points are far apart. For general $z_-$ and $z_+$ one can use a uniform approximation involving parabolic cylinder functions, finding \cite{PhysRev.91.174,Berry:1972na}
\begin{align}\label{connectioniml}
\psi(z)&=\frac{C_-}{\sqrt{|\kappa(z)|}}e^{-p\int_{0}^{z}dz'\,|\kappa(z')|}+\frac{C_+}{\sqrt{|\kappa(z)|}}e^{p\int_{0}^{z}dz'\,|\kappa(z')|},\hspace{10 mm} z<z_-,
\end{align}
where the connection coefficients are given by 
\begin{align}
C_+&=\sin(S(z_-,z_+))e^{-S(0,z_-)}\\
C_-&=\sqrt{\frac{2}{\pi}}\left(\frac{\pi e}{S(z_-,z_+)}\right)^{S(z_-,z_+)/\pi}\Gamma\left(\frac{S(z_-,z_+)}{\pi}+\frac{1}{2}\right)\cos(S(z_-,z_+))e^{S(0,z_-)}.
\end{align}
In the limit of large $S(z_-,z_+)$, this reduces to the answer obtained by applying the WKB connection formulae at $z_+$ and $z_-$ successively.\\
\indent Finally, we must match (\ref{connectioniml}) to the correct boundary asymptotics (\ref{boundarybehavior}) as above. In terms of modified Bessel functions, we have\footnote{The asymptotic expansion of modified Bessel functions at large argument is subtle, due to the presence of a Stokes line on the real axis. We refer the reader to Chapter 2 of \cite{dingle} for the derivation of the asymptotic expansion of (\ref{modifiedbessel}). For half-integer $\nu$, the modified Bessel functions are elementary and it is simple to check that (\ref{modifiedbessel}) has the correct asymptotic behavior.}
\begin{align}\label{modifiedbessel}
\psi(z)&=C_+\sqrt{\frac{\pi pz}{2}}(I_{-\nu}(p\sqrt{u^2-1}z)+I_\nu(p\sqrt{u^2-1}z))\notag\\
&\hspace{5 mm}+C_- \sqrt{\frac{2pz}{\pi}}K_\nu(p\sqrt{u^2-1}z),\hspace{5 mm}z\sim 1/p.
\end{align}
The retarded Green's function (\ref{grdef}) is
\begin{align}
G_R(\omega,\ell)=\frac{\Gamma(-\nu)}{\Gamma(\nu)}\left(\frac{p^2-\omega^2}{4}\right)^{\nu}\frac{C_--C_+\sin(\pi\nu)}{C_-+C_+\sin(\pi\nu)}.
\label{eq:GRimspbh}
\end{align}
\indent In the limit of large $S(0,z_-)$, the poles of $G_R$ are located at
\begin{align}\label{actionqnmiml}
S_m(z_-,z_+)=\pi\left(m+\frac{1}{2}\right)+\sqrt{\frac{\pi}{2}}\left(\frac{m+\frac{1}{2}}{e}\right)^{m+\frac{1}{2}}\frac{\sin(\pi\nu)}{m!}e^{-2S_m(0,z_-)},\hspace{5 mm}m=0,1,\ldots,
\end{align}
with both polynomial corrections as well as corrections of the form $e^{-4 S_m(0,z_-)}$, $e^{-6 S_m(0,z_-)}$, $...$ \ . We will see in the next section that including the latter leads to a correct prediction of the locations of the bouncing singularities depicted in Figures \ref{fig:1bouncefig} and \ref{fig:2bouncefig}. However, we were not able to match the shapes and heights of the bounces using exponentially small corrections to \eqref{actionqnmiml}, so these corrections should not be trusted.

The residues are
\begin{align}\label{residuesiml}
 \underset{\hspace{2mm}\omega\to \omega_m}{\text{Res}}G_R(\omega,\ell)=-\frac{1}{\nu\Gamma(\nu)^2}\left(\frac{(\ell+\alpha)^2-\omega_m^2}{4}\right)^{\nu}\frac{\sqrt{2\pi^3}}{m!}\left(\frac{m+\frac{1}{2}}{e}\right)^{m+\frac{1}{2}}\frac{e^{-2S_m(0,z_-)}}{\frac{dS_m(z_-,z_+)}{d\omega}}.
\end{align}
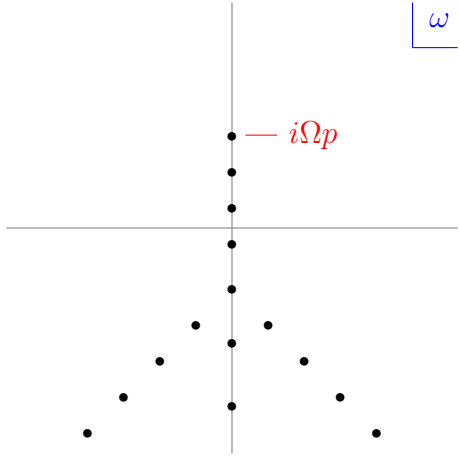
\begin{figure}
\centering
\begin{tikzpicture}[scale=1.2]
\draw[gray] (-2.5,0) -- (2.5,0);
\draw[gray] (0,-2.5) -- (0,2.5);
\foreach \Point in {(0,1),(0,.6),(0,.2),(0,-.2),(0,-.7),(0,-1.3),(0,-2),(.4,-1.1),(.8,-1.5),(1.2,-1.9),(1.6,-2.3),(-.4,-1.1),(-.8,-1.5),(-1.2,-1.9),(-1.6,-2.3)}{
    \node at \Point {\textbullet};}
                   \node[blue] at (2.3,2.3)  {$\omega$};
                      \draw[blue] (2.5,2) -- (2,2);
      \draw[blue] (2,2) -- (2,2.5);
         \draw[red] (.15,1.03) -- (.5,1.03);
          \node[red] at (.9,1.03)  {$i\Omega p$};
\end{tikzpicture}
 \caption{The spectrum of quasi-normal modes for $\ell+\alpha=ip$ imaginary. There is a series of imaginary modes starting in the upper half plane at $\omega=i\Omega p$, as well as two lines of complex modes in the lower half plane.  \label{imaginaryellqnms}}
\end{figure}
\indent Now let us analyze the spectrum for $m\ll p$, which consists of modes localized near the photon sphere. The action integral reduces to 
\begin{align}
S(z_-,z_+)=\frac{\pi p}{\sqrt{2(V^{\ell+\alpha=ip}_{\text{eik}})''(z_{\text{phot}})}}(-V^{\ell+\alpha=ip}_{\text{eik}}(z_\text{phot})-u^2),
\end{align}
where $z_{\text{phot}}$ is the location of the photon sphere. Solving (\ref{actionqnmiml}) at large $p$, we find
\begin{align}\label{eikonaliml}
\omega_m=\Omega (\ell+\alpha)-i\gamma\left(m+\frac{1}{2}\right),\hspace{10 mm}m\ll |\ell|,
\end{align}
where we have defined the Lyapunov exponent
\begin{align}\label{eq:LyapunovDef}
\gamma\equiv \sqrt{-\frac{(V_{\text{eik}}^{\ell+\alpha=ip})''(z_{\text{phot}})}{2V^{\ell+\alpha=ip}_{\text{eik}}(z_{\text{phot}})}} = \sqrt{d-2} \ \Omega .
\end{align}
In Table \ref{tab:n5}, we compare the analytic prediction (\ref{eikonaliml}) to numerics, finding agreement to a high level of accuracy. This spectrum of bound states is displayed in Figure \ref{imaginaryellqnms}, along with the rest of the quasi-normal modes in the lower half plane. 
\begin{table}
\centering
\begin{tabular}{ |c|c|c| } 
\hline
 m & $\mathtt{QNMSpectral}$ & Eikonal  \\ 
 \hline
 0 & 222.79 $i$ & 222.82 $i$  \\ 
 1 & 221.24 $i$ & 221.23 $i$  \\ 
 2 & 219.70 $i$ & 219.65 $i$  \\ 
 3 & 218.18 $i$ & 218.07 $i$  \\ 
 4 & 216.69 $i$ & 216.49 $i$  \\ 
\hline
\end{tabular}
\caption{We compare numerical values for the photon sphere QNMs obtained with the Mathematica package $\mathtt{QNMSpectral}$ \cite{Jansen:2017oag} against the eikonal prediction \eqref{eikonaliml}.  Here $d=4$, $p=200$, $\nu=2$ and $\mu=1$.}
\label{tab:n5}
\end{table}

\subsection{The flat space limit}
\label{sec:flatspace}
\indent It is instructive to consider the limit $\mu\ll1$, which describes a small black hole in AdS. For real $\ell$, the poles and residues can be computed order by order in $\mu$ using (\ref{polecondition}) and (\ref{resreall}). For example, in $d=4$ we get 
\begin{align}
\text{Re }\omega_m&=\omega_m^{(0)}-\mu \frac{3m^2}{\ell}+\ldots \\
\frac{ \underset{\hspace{2mm}\omega\to \omega_m}{\text{Res}}G_R(\omega,\ell)}{ \underset{\hspace{2mm}\omega\to \omega_m}{\text{Res}}G^{(0)}_R(\omega,\ell)}&=1-\mu\frac{3m((\nu+2) \ell+2m(\nu+1))}{2\ell(m+\ell)}+\ldots,
\end{align}
where $\omega_m^{(0)}$ and $\text{Res}_{\omega\to\omega_m}G^{(0)}_R(\omega,\ell)$ are given by (\ref{omegan0}) and (\ref{res0}). On the boundary, these quasi-normal modes correspond to the heavy-light double-twist operators $[HL]_{m\ell }$ in the Regge limit $m,\ell\to \infty$ with $m/\ell$ fixed, see e.g.\ \cite{Kulaxizi:2018dxo,Karlsson:2019qfi,Fitzpatrick:2019efk,Li:2019zba,Karlsson:2019txu,Li:2020dqm,Parnachev:2020zbr,Dodelson:2022eiz} for related work. The mode energies and residues match the double-twist spectrum and OPE coefficients $c^{[HL]}_{HL}$ obtained by alternative methods. The crucial point here is that these modes are localized near the AdS boundary, not near the black hole photon sphere. Therefore, in the flat space limit we do not recover the eikonal spectrum of asymptotically flat black holes.  \\
\indent For imaginary spin the situation is different. As explained above, the highest bound states on the imaginary axis correspond to modes localized near the photon sphere. Therefore the eikonal spectrum (\ref{eikonaliml}) for imaginary $\ell+\alpha$ smoothly approaches the eikonal spectrum in an asymptotically flat black hole in the limit $\mu\to 0$. Let us work in the case $d=3$, which corresponds to a small black hole in four dimensions. The spectrum becomes
\begin{align}
\omega_m=\frac{2}{3\sqrt{3}r_s} \left(\ell+\frac{1}{2}\right)-\frac{2i}{3\sqrt{3}r_s}\left(m+\frac{1}{2}\right),\hspace{10 mm}m\ll|\ell|,r_s\ll L_{AdS},
\end{align}
which matches the standard asymptotically flat results \cite{Goebel:1972,Ferrari:1984zz,Mashhoon:1985,Schutz:1985km,Iyer:1986np,Berti:2009kk}, see Appendix \ref{sec:Signatures} for a review.

\section{The bulk-cone singularity from thermal Regge poles}\label{bulkcone}
In the previous section we studied the spectrum of QNMs for both real and imaginary spin, and found that in the case of imaginary spin the spectrum is controlled by the photon sphere. In fact, the same is true for the thermal Regge poles at large imaginary frequency. In this section, we will consider the position space correlator in special kinematics where the two boundary points are connected by a light ray in the bulk \cite{Hubeny:2006yu,Dodelson:2020lal}. As reviewed in the introduction, we expect a singularity in this kinematic configuration, and the role of the Regge poles will be to precisely reproduce this singularity.
\subsection{Position space and complex spin}
Let us first generalize the Regge expansion of the correlator from Section \ref{Sec:ReggeInfVol} to thermal correlators on the sphere $S^{d-1}$, whose radius we set to $R=1$. Recall that the Wightman function is obtained from the Euclidean correlation function as follows,
\be
G_W(t,\theta) =\lim_{\epsilon\to 0} G_E(\tau = \epsilon +i t,\theta) . 
\ee

On the other hand, the Euclidean correlator on $S^{d-1}$ admits the following representation\footnote{The relationship between $G_E$ and $G_R$ could be modified at $\omega=0$, see e.g. \cite{Meyer:2011gj}. This subtlety is not relevant in exploring the singularities of the Lorentzian correlators so we ignore it. } 
\be
G_E(\tau,\theta) = \frac{1}{\beta}\sum_{n=-\infty}^{\infty} e^{i \zeta_n \tau} \sum_{{\ell}=0}^{\infty}G_{R}\left(\omega = i |\zeta_n|, \ell\right) \frac{\ell+\alpha}{\alpha}C_\ell^{(\alpha)}(\cos\theta),
\ee
where $\alpha=\frac{d-2}{2}$, $C_\ell^{(\alpha)}(\cos\theta)$ are the Gegenbauer polynomials, and the Matsubara frequencies $\zeta_n = \frac{2 \pi n}{\beta}$ were introduced in (\ref{matsubara}). We have again used the relationship between the Euclidean and retarded two-point functions in momentum space. In order to analytically continue in spin $\ell$, we first note that $C_\ell^{(\alpha)}(\cos\theta)$ vanishes on the negative integers between $3-d,\ldots,-1$, so that we can trivially extend the summation to start at $\ell=-\lfloor\alpha\rfloor$. We now use a Sommerfeld-Watson transform to rewrite the sum as an integral
\be\label{eq:SW}
G_E(\tau,\theta) = \frac{i}{2\beta}\sum_{n=-\infty}^{\infty} e^{i \zeta_n \tau} \int_\gamma \frac{d\ell}{\sin\pi\ell}\,G_{R}\left(i|\zeta_n|, \ell\right) \frac{\ell+\alpha}{\alpha}C_\ell^{(\alpha)}(-z),
\ee 
where $\gamma$ is the sum of contours running clockwise around the integers $-\lfloor\alpha\rfloor,\ldots,\infty$ and we introduced $z \equiv \cos \theta$. 

\begin{figure}
    \centering
    \begin{tikzpicture}[
  arrowdecor/.style={
    postaction={decorate,decoration={markings,mark=at position 0.5 with {\arrow{<}}}}
  }]

  \draw (-5.5,0) -- (5.5,0) node[right]{${\rm Re}(\ell)$};
  \draw (0,-3.5) -- (0,3.5) node[above]{${\rm Im}(\ell)$};

  \foreach \x in {-5,...,5}
    \fill (\x,0) circle (0.05);

  \foreach \x in {-5,...,5}
    \draw (\x,0) circle (0.15);
    
  \foreach \x in {-5,...,5}
    \draw[<-,red] (\x+0.15,0) arc(0:360:0.15);
    
  \draw[thick, blue, ->] (-5.15,0.3) -- (5.15,0.3);
  \node[above,blue] at (5.15,0.3) {$\mathcal{C}_+$};
  
  \draw[thick, blue,<-] (-5.15,-0.3) -- (5.15,-0.3);
  \node[below,blue] at (-5.15,-0.3) {$\mathcal{C}_-$};

  \foreach \x in {1,2,3}
    \node[above] at (-1,\x) {$\times$};
    \foreach \x in {-1,-2,-3}
    \node[below] at (-1,\x) {$\times$};

\end{tikzpicture}
    \caption{We start with a sum over integer spins indicated by the contours in red, which we deform to the contours $\mathcal{C}_+$ and $\mathcal{C}_-$. Because of the symmetry properties around $\ell=\frac{2-d}{2}$, the contributions from $\mathcal{C}_+$ and $\mathcal{C}_-$ are equal. We can then close the contour $\mathcal{C}_+$ in the upper half plane, picking up the Regge poles. The red contours around the integers come in three different classes. For $\ell\geq0$ these are the spins we started with, for $\ell=3-d,\ldots,-1$ the Gegenbauer polynomials vanish and give no contribution, and finally for $\ell\leq 2-d$ we get dual poles due to the symmetry of the integrand around $\ell=\frac{2-d}{2}$. In this figure we have assumed that there are no Regge poles on the real axis, see Appendix \ref{reggebounds}. In Figure \ref{fig:reggepoles} we computed Regge poles numerically using the Mathematica package $\mathtt{QNMSpectral}$ \cite{Jansen:2017oag}.}
    \label{fig:ReggePlane}
\end{figure}
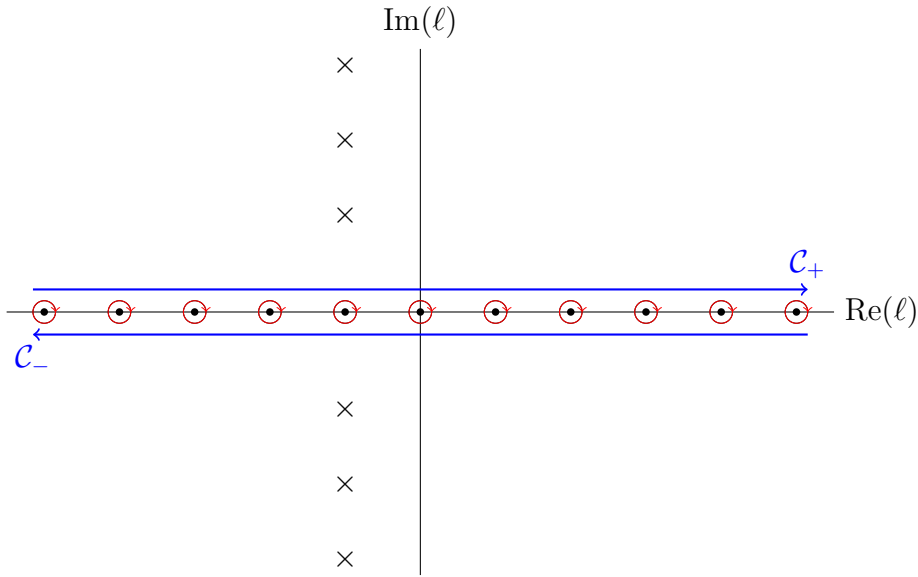

Next, notice that the wave equation depends on $\ell$ only through $\ell(\ell+2\alpha)$. It follows that $G_R(\omega,\ell)$ is analytic in $\ell$ and
\be 
    G_R(\omega,k-\alpha)=G_R(\omega,-k-\alpha),
\ee
where $\ell\equiv k-\alpha$. Meanwhile, the Gegenbauer polynomials satisfy the relation 
\begin{align}
C_{k-\alpha}^{(\alpha)}(-z)=(-1)^{2\alpha+1}C_{-k-\alpha}^{(\alpha)}(-z),
\end{align}
so the full integrand in \eqref{eq:SW} is even around $k=-\alpha$. This allows us to extend the integration along the whole real axis as in Figure \ref{fig:ReggePlane},
\be
G_E(\tau,\theta) = \frac{i}{4\alpha\beta}\sum_{n=-\infty}^{\infty} e^{i \zeta_n \tau} \int_{\mathcal{C}_++\mathcal{C}_-} \frac{k\, dk}{\sin\left(\pi(k-\alpha)\right)}G_{R}\left(i |\zeta_n|, k-\alpha\right) C_{k-\alpha}^{(\alpha)}(-z),
\ee 
where $\mathcal{C}_+$ runs just above the real axis to the right and $\mathcal{C}_-$ runs just below the real axis to the left. Due to the symmetry under $k\to-k$, this can be written as an integral only over $\mathcal{C}_+$ above the real axis. \\
\indent We then deform the contour in the upper half plane, picking up all the Regge poles with ${\rm Im}\, k>0$. Here we used the fact that the Gegenbauer polynomials have the following leading behavior at large imaginary $k$,
\be \label{gegasymp}
   C_{i p-\alpha}^{(\alpha)}(-z)=-{ie^{i \pi \alpha}p^{\alpha-1} \over (2 \sin\theta)^\alpha \Gamma(\alpha)}e^{p(\pi-\theta)}  \Big( 1 + \mathcal{O}\left(\frac{1}{p}\right) \Big) , ~~~ 0<\theta<\pi.
\ee
The arc therefore vanishes as long as $\theta\leq\pi$ thanks to the $1/ \sin (\pi (k-\alpha))$ factor. This leads to the Regge expansion of the Euclidean correlator,
\be \label{reggefinitevolume}
G_E(\tau,\theta) = -\frac{\pi}{\alpha\beta}\sum_{n=-\infty}^{\infty}\sum_m e^{i \zeta_n \tau}\frac{k_{mn} C_{k_{mn}-\alpha}^{(\alpha)}(-z)}{\sin\left(\pi(k_{mn}-\alpha)\right)} \underset{\hspace{2mm}k\to k_{mn}}{\text{Res}}G_{R}\left(i |\zeta_n|,k-\alpha\right),
\ee 
where $k_{mn}\equiv k_m(i|\zeta_n|)$ are the Regge poles in the upper half plane.

In Appendix \ref{app:derivingwindings} we show that the formula \eqref{reggefinitevolume} can be rewritten in the following form
\be
\label{eq:gensumformula}
G_E(\tau, \theta) = \sum_{j=0}^\infty \Big( g_E(\tau, |\theta| + 2 \pi j ) +(-1)^{2\alpha} g_E(\tau, 2 \pi - |\theta| + 2 \pi j ) \Big) \ , ~~~ | \theta| < \pi , 
\ee
where
\be\label{littlege}
g_E(\tau,\theta) &= {4^{1-\alpha} \over (\sin \theta)^{2 \alpha-1}}\frac{\pi}{\alpha\beta}\sum_{n=-\infty}^{\infty} e^{i \zeta_n \tau} \sum_m \underset{\hspace{2mm}k\to k_{mn}}{\text{Res}}G_{R}\left(i |\zeta_n|, k-\alpha\right) e^{i (1+k_{mn}-\alpha) \theta} \nn \\ 
&\hspace{5mm}\times {\Gamma(k_{mn}+\alpha) \over \Gamma(k_{mn})\Gamma(\alpha)} \ _2 F_1 (1-\alpha,1+k_{mn}-\alpha,1+k_{mn}, e^{2 i\theta}).
\ee
We will see that the sum over $j$ in (\ref{eq:gensumformula}) corresponds to summing over the winding number of bulk geodesics. Note that the formula \eqref{eq:gensumformula} is manifestly KMS invariant. It is also invariant under $\theta \to - \theta$. For later reference we note the asymptotics of the hypergeometric function at large imaginary spin,
\begin{align}\label{hyperasymp}
&{\Gamma(i p+\alpha) \over \Gamma(i p )\Gamma(\alpha)} \ _2 F_1 (1-\alpha,1+i p -\alpha,1+ i p , e^{2 i \theta}) \nn \\ 
&\simeq e^{{i \pi \alpha\over 2} } {p^{\alpha} \over \Gamma(\alpha)} (1-e^{2 i \theta})^{\alpha-1} \Big( 1  +\mathcal{O}\left(\frac{1}{p}\right) \Big) . 
\end{align}

\subsection{The ordinary light-cone}
First, let us understand how the usual light-cone emerges from the Regge expansion (\ref{reggefinitevolume}). In analogy to the infinite volume case, it is natural to expect that the Regge poles that reproduce the ordinary light cone are purely imaginary, $k_m(i|\zeta_n|)=ip_m(i|\zeta_n|)$, with $p_m(i|\zeta_n|)>|\zeta_n|$. Indeed, the modes (\ref{finitevolumeordinary}) with one turning point are precisely of this form. 
\\
\indent Let us now analyze the contribution of the modes (\ref{finitevolumeordinary}) to the function $g_E(\tau,\theta)$ defined in (\ref{littlege}). The residues of these modes are given by (\ref{finitevolumeordinaryres}). For $\tau,\theta\ll 1$, we can approximate the sums over Regge poles and Matsubara frequencies by integrals. Using the asymptotics (\ref{gegasymp}), the kernel in the integrand takes the following simple form at large $p$ up to power-law corrections,
\begin{align}
\label{eq:largespingegenSimple}
\frac{p C^{\alpha}_{ip-\alpha}(-z)}{\sinh(\pi(p+i\alpha))}={-2ip^{\alpha} e^{-p \theta} \over (2 \sin\theta)^\alpha \Gamma(\alpha)} \left(1 + \mathcal{O}\left(\frac{1}{p}\right)\right) . 
\end{align}
where we only kept the leading large $p$ asymptotic. For $\tau,\theta\ll1$, the expansion (\ref{reggefinitevolume}) then becomes 
\begin{align}
\label{eq:spherelightcone}
G_E(\tau,\theta)=\frac{1}{2^{2\nu+\alpha}\nu\Gamma(\nu)^2\Gamma(\alpha+1)\theta^\alpha}\int_{0}^{\infty}dp\, \int_{-p}^{p} d\zeta\, e^{i\zeta \tau}p^\alpha e^{-p\theta}(p^2-\zeta^2)^{\nu}.
\end{align}

Performing the integrals in \eqref{eq:spherelightcone}, we get for the leading light-cone asymptotic
\be\label{leadinglc}
G_W(t,\theta) \simeq {2 \Gamma(\Delta) \over \Gamma\left(\frac{d}{2}\right) \Gamma\left(\Delta-\frac{d}{2}\right)} {1 \over 2^\Delta \theta^\Delta (\theta-t)^\Delta} , ~~~0 < \theta \ll 1 ,
\ee
where the computation above only matches the leading $\theta \ll 1$ asymptotic of the leading light-cone singularity ${1 \over (\theta-t)^\Delta} {1 \over \theta^{\Delta}}$. We get the same normalization as the one obtained for real spins after \eqref{res0}. Going beyond that requires taking into account ${1 \over \ell}$
corrections both in \eqref{finitevolumeordinaryres} and in \eqref{eq:largespingegenSimple}. It would be interesting to check explicitly that the subleading terms in the small $\theta$ expansion come out correctly.
\subsection{Bulk-cone singularities}

Now that we have understood the ordinary light-cone, we can address the new singularity on the bulk light cone. Since the bulk light cone is controlled by the photon sphere at late times, we expect that the Regge poles near the bottom of the potential in Figure \ref{potentialiml} are responsible for reproducing the late time  bulk-cone. Let us now confirm this expectation.\\
\indent We consider the contribution of the modes (\ref{actionqnmiml}) with $\ell+\alpha=ip$ and $p_m(\zeta_n)<|\zeta_n|<\Omega p_m(\zeta_n)$ to the Regge expansion (\ref{reggefinitevolume}). The residues are given by (\ref{residuesiml}), with the factor of $dS(z_-,z_+)/d\omega$ in the denominator replaced by $dS(z_-,z_+)/dk$, since we are considering Regge poles, not quasi-normal modes. Replacing the sums by integrals as usual and using the asymptotic formula (\ref{hyperasymp}) then gives 
\begin{align}
g_E(\tau,\theta)&=\frac{e^{i\pi\alpha\left\lfloor\frac{\theta}{\pi}\right\rfloor}}{2^{2\nu+\alpha}\nu\Gamma(\nu)^2\Gamma(\alpha+1)|\sin\theta|^\alpha}\notag\\
&\hspace{10 mm}\times\int_{0}^{\infty}dp\, \int_{p}^{\Omega p} d\zeta\, e^{-i\zeta \tau-p\theta}p^\alpha(\zeta^2-p^2)^{\nu}e^{-2S(0,z_-)},
\end{align}
where $\lfloor x\rfloor$ is the floor of $x$, i.e. the largest integer $n$ with $n\le x$. In this formula we have assumed that $S(z_-,z_+)\gg1$. We will check this assumption later. We have also neglected the contribution from $\zeta<0$ to the integral, since $e^{i\zeta \tau}$ can never produce a singularity under Wick rotation $\tau\to it$.\\
\indent Next let us perform the integral over $p$, for which it is useful to define $u=\zeta/p$. Recall that the action integral can be written as
\begin{align}\label{actionintegral}
2S(0,z_-)=puT(u)-p\Theta(u)\ge 0,
\end{align}
where $T(u)$ and $\Theta(u)$ are twice the elapsed time and angle of a null geodesic from the boundary to $z_-$, 
\begin{align}
T(u)&=2\int_{r(z_-)}^{\infty}\frac{dr}{f(r)}\frac{u}{\sqrt{u^2-\frac{f(r)}{r^2}}} \ ,  \label{tofu}\\
\Theta(u)&=2\int_{r(z_-)}^{\infty}\frac{dr}{r^2}\frac{1}{\sqrt{u^2-\frac{f(r)}{r^2}}}.
\label{thofu}
\end{align}
The result of the $p$ integral is then
\begin{align}
g_E(it+\epsilon,\theta)&=\frac{\Gamma(2\nu+\alpha+2)e^{i\pi\alpha\left\lfloor\frac{\theta}{\pi}\right\rfloor}}{2^{2\nu+\alpha}\nu\Gamma(\nu)^2\Gamma(\alpha+1)|\sin\theta|^\alpha}\notag\\
&\hspace{10 mm}\times\int_{1}^{\Omega } du\, \frac{(u^2-1)^{\nu}}{(\theta-\Theta(u)-u(t-T(u))+i\epsilon)^{2\nu+\alpha+2}}.
\end{align}
\indent Note that the integral over $p$ converges when 
\be
t-\theta<0 ,
\ee
or in other words in the spacelike region. From there we can continue the integral to $t>\theta$ using the $i \epsilon$ prescription. We then have a candidate singularity at 
\be
u_* t = u_* T(u_*) - \Theta(u_*) + \theta . 
\ee
However for it to be an actual singularity the integration contour must be pinched, which happens for
\be
\partial_u (u T(u) - \Theta(u)) |_{u = u_*} = t . 
\ee
Using the fact that $u T'(u) - \Theta'(u) = 0$, we thus find that the condition for the pinch becomes
\be
t = T(u) \geq \pi , \nn \\
\theta = \Theta(u) \geq \pi . 
\ee
\indent Now let us compute the functional form of the correlator near the singularity. To do so, we expand around the pinch singularity as $u=u(t)+\delta u$, where $u(t)$ is the solution to $T(u(t))=t$. Expanding to second order gives
\begin{align}
\theta-\Theta(u)-u(t-T(u))=\theta-\Theta(u(t))+\frac{T'(u(t))}{2}\delta u^2.
\end{align}
The $\delta u$ integral can then be performed as follows,
\begin{align}
\int_{-\infty}^{\infty}  \frac{d\delta u}{\left(\theta-\Theta(u(t))+\frac{T'(u(t))}{2}\delta u^2+i\epsilon\right)^{2\nu+\alpha+2}}&=\frac{\Gamma\left(2\nu+\alpha+\frac{3}{2}\right)}{\Gamma(2\nu+\alpha+2)}\sqrt{\frac{2\pi}{T'(u(t))}}\notag\\
&\hspace{5 mm}\times \frac{1}{(\theta-\Theta(u(t))+i\epsilon)^{2\nu+\alpha+\frac{3}{2}}}.
\end{align}

Close to the singularity we have $\theta-\Theta(u(t)) = u(t) (t_{\text{BC}}(\theta)-t)$, where $t_{\text{BC}}(\theta) \equiv T(u(t))|_{\Theta(u(t))=\theta}$ is defined as the time it takes a geodesic emanating from the boundary to return to the boundary after traversing an angle $\theta$. In this way we get for the singularity
\begin{align}
g_E(it+\epsilon,\theta)&=\frac{\sqrt{2\pi}\Gamma\left(2\nu+\alpha+\frac{3}{2}\right)(u(t)^2-1)^{\nu} (u(t))^{-(2\nu+\alpha+\frac{3}{2})}}{2^{2\nu+\alpha}\nu\Gamma(\nu)^2\Gamma(\alpha+1)|\sin\theta|^\alpha\sqrt{T'(u(t))} }\frac{e^{i\pi\alpha\left\lfloor\frac{\theta}{\pi}\right\rfloor}}{(t_{\text{BC}}(\theta) - t +i\epsilon)^{2\nu+\alpha+\frac{3}{2}}}.
\end{align}
\indent The singularities of the Wightman function can now be obtained by summing $g_E$ over the winding number $j$ as in (\ref{eq:gensumformula}). Taking $0<\theta<\pi$, we are left with the final result
\be\label{eq:GWleadBC}
G_W(t,\theta)&\simeq {2 \Gamma(\Delta) \over \Gamma(d/2) \Gamma(\Delta-d/2)} \times \frac{2^{\frac{d}{2}-2 \Delta }
   \Gamma \left(2 \Delta-\frac{d-1}{2}\right)}{\Gamma (\Delta ) \Gamma
   \left(\Delta -\frac{d-2}{2}\right)} {\sqrt{2 \pi} \over ( \sin \theta)^{{d-2 \over 2}}} \\
&\hspace{10 mm}\times\frac{(u(t)^2-1)^{\Delta -d/2}  }{(u(t))^{2 \Delta -{d-1 \over 2}} \sqrt{T'(u(t))}} \sum_{j=1}^\infty (-1)^{jd} \Big(X_j^+(t,\theta) + e^{i \pi {d-2 \over 2}} X_j^-(t,\theta) \Big)\notag,
\ee
where
\be
X_j^\pm(t,\theta) = {1 \over \Big(t_{\text{BC}}(2 \pi j  \pm \theta) - t + i \epsilon \Big)^{2 \Delta -{d-1 \over 2}}} . 
\ee
Close to the singularity the following conditions are satisfied,
\be\label{conditionsnobounce}
t &= T(u) , \nn \\
2 \pi j \pm \theta &=\Theta(u) . 
\ee
The solutions to these conditions satisfy $1 \leq u(t) \leq \Omega$, where $u(t) \to \Omega$ corresponds to the late time limit $t \to \infty$. Let us remind the reader that the formula above is written in the normalization where the unit operator contributes to the OPE as follows, $ {2 \Gamma(\Delta) \over \Gamma(d/2) \Gamma(\Delta-d/2)} {1 \over 2^\Delta (\cos t -\cos \theta)^\Delta}$.  The pre-factor $\frac{2^{d/2-2 \Delta } \Gamma \left(2 \Delta-\frac{d-1}{2}\right)}{\Gamma (\Delta ) \Gamma
   \left(\Delta -\frac{d-2}{2}\right)}$ is $\mathcal{O}(1)$ for fixed $d$ and any $\Delta$.\\
   \indent Let us now comment on the singularities corresponding to null geodesics that bounce off the AdS boundary. These were predicted in \cite{Dodelson:2020lal}, and are depicted in Figures \ref{fig:1bouncefig} and \ref{fig:2bouncefig}. As mentioned above, the pole condition (\ref{actionqnmiml}) receives exponentially small corrections, which leads to corrections to the residues (\ref{residuesiml}) that are proportional to $e^{-2nS(0,z_-)}$ with $n>1$. Repeating the analysis leading to (\ref{eq:GWleadBC}) gives new singularities at 
   \begin{align}\label{eq:BounceDef}
       t&=nT(u)\notag\\
       2\pi j\pm \theta&=n\Theta(u).
   \end{align}
   This is indeed the expected location of the singularity with $n-1$ bounces, and we will numerically confirm the presence of these singularities in the next section. However, we were not able to match the predictions for the shape and size of the bounces to numerics, so we omit the details of the computation. It would be very interesting to understand why the naive calculation fails; perhaps more sophisticated WKB techniques are required.
\subsection{Corrections}

There are two types of corrections to the singularity structure above that we neglected. The first come from powers ${1 \over p}$, which enter both the expansion of the Gegenbauer polynomials and the residues. These will produce subleading singularities.  \\
\indent Second, we approximated the sums over $\zeta$ and $p$ by integrals. By the Euler-Maclaurin formula, this is equivalent to neglecting contributions from the endpoints at $\zeta=p+c_1$ and $\zeta=\Omega p-c_2$, where $c_1$ and $c_2$ are order one constants. The leading endpoint contribution from $\zeta=p+c_1$ yields a subleading singularity on the ordinary light-cone,
\begin{align}
G_E(it+\epsilon,\theta)&\propto \int_{0}^{\infty}dp\, e^{p(t-\theta-i\epsilon)}p^{\alpha+\nu} \propto \frac{1}{(\theta-t+i\epsilon)^{\Delta}} \ .
\end{align}
At small $\theta$, this is suppressed compared to the light-cone asymptotic (\ref{leadinglc}) by a power $\theta^\Delta$, so it can be neglected. Further endpoint corrections involve the derivative of the integrand at the endpoints, and it is straightforward to check that these are subleading as well. \\
\indent Similarly, the endpoint at $\zeta=\Omega p-c_2$ leads to a singularity of the form $\frac{1}{(2\pi j\pm \theta-\Omega t+i\epsilon)^{\Delta}}$. However, $2\pi j\pm \theta=\Omega t$ is never satisfied when the saddle point conditions (\ref{conditionsnobounce}) hold, so this singularity is fictitious and can be discarded. We conclude that the endpoint contributions to the bulk-cone singularity are suppressed away from $t = \pi$ and $t=\infty$.

\subsection{Dependence of singularities on spacetime dimension}
\label{sec:dimdep}

One interesting consequence of (\ref{eq:GWleadBC}) is that the structure of the singularities depends on the number of spacetime dimensions. We now classify the possible cases. Recall that $\alpha = {d - 2 \over 2}$.

Consider first the case when $d \in 4 \mathbb{Z}_{>0} + 2 = 6,10,...$. In this case $e^{i \pi \alpha} = 1$ and all singularities enter in the same way as $\sum_k (X_k^{+} + X_k^-)$. We can say that in this case the structure of singularities is \emph{one-fold}.

The next simplest case is $d \in 4 \mathbb{Z}_{>0} = 4,8,12,...$. In this case $e^{i \pi \alpha} = -1$, and the singularities enter with alternating signs, $\sum_k (X_k^{+} - X_k^-)$. The structure of singularities is therefore \emph{two-fold}.

Let us next take the number of dimensions to be odd. In this case the structure of singularities is \emph{four-fold}, as was first pointed out in the case of asymptotically flat black holes in $d=3$ \cite{Dolan:2011fh}. For $d=4 \mathbb{Z}_{>0} - 1 = 3 , 7, 11, ... $ we have $e^{i \pi \alpha} = i$. The singularity structure then takes the form $\sum_{j} (-1)^j (X_j^{+} + i X_j^-)$.
Finally, for $d=4 \mathbb{Z}_{>0} + 1 = 5 , 9, 13, ... $ we have $e^{i \pi \alpha} = -i$. The singularity structure then takes the form $\sum_{j} (-1)^j (X_j^{+} - i X_j^-)$.

It is interesting to analyze the implications of this dimension-dependent structure for the retarded two-point function, which is defined as
\be
\label{eq:relRandW}
G_R(t, \theta) = i \theta(t) \Big( \la {\cal O}(t,\theta) {\cal O}(0)\ra - \la {\cal O}(0) {\cal O}(t,\theta) \ra \Big) = -2\theta(t)\text{Im }G_W(t, \theta) .
\ee
In the last step we used that for real operators $\la {\cal O}(0) {\cal O}(t,\theta) \ra = \la {\cal O}(t,\theta) {\cal O}(0)\ra^*$. This formula has  interesting consequences for the structure of singularities when $2 \Delta + {1-d \over 2}$ is a positive integer. The reason is that in this case $X_k-X_k^*$ is a delta-function or its derivatives, whereas $i (X_k+X_k^*)$ is a power-like singularity. A simple example of this type is a massless perturbation with $\Delta= d$ in odd $d$. In the next section we will study this example numerically in $d=3$ and confirm the results discussed here.
\subsection{Thermalization of the bulk cone}
\begin{figure}
     \centering
     \begin{subfigure}[b]{0.45\textwidth}
         \centering
    \includegraphics[width=\textwidth]{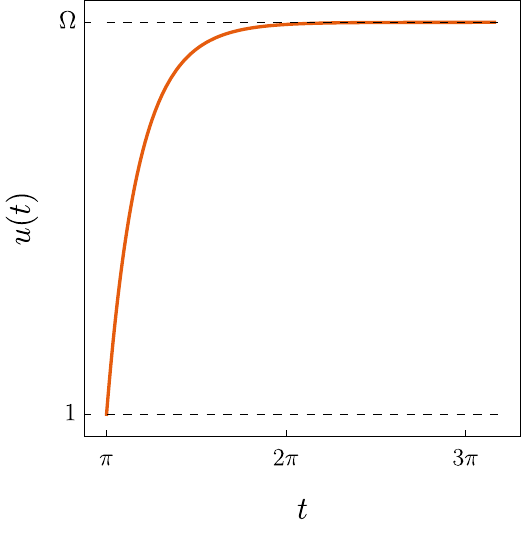}
         \caption{}
         \label{fig:u(t)}
     \end{subfigure}
     \hfill
     \begin{subfigure}[b]{0.42\textwidth}
         \centering
    \includegraphics[width=\textwidth]{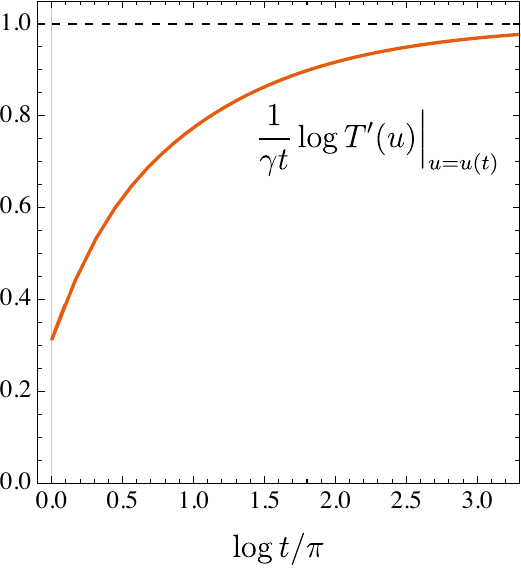}
         \caption{}
         \label{fig:LogT}
     \end{subfigure}
     \caption{In (a) we plot $u(t)$ in $d=4$ for $\mu=1$. The dashed lines correspond to $u_{\text{min}}=1$ and $u_{\text{max}}=\Omega$. At late times $u(t)$ is seen to approach $\Omega$. In (b) we plot $\frac{1}{\gamma t}\log T'(u) \Big|_{u=u(t)}$ which approaches $1$ ($\gamma=\sqrt{2}\Omega\approx 1.58$ in $d=4$ and $\mu=1$). Equivalently, $T'(u) \sim e^{\gamma t}$ at late times.}
     \label{fig:Tandu}
\end{figure}
\indent The formula (\ref{eq:GWleadBC}) for the leading bulk-cone singularity reveals a curious fact: the bulk-cone singularity behaves as $(t_{\text{BC}}(\theta)-t)^{-2\Delta+\frac{d-1}{2}}$. While this power is weaker than the Euclidean singularity $(\tau^2+\theta^2)^{-\Delta}$, it is actually stronger than the ordinary light-cone singularity $(\theta-t)^{-\Delta}$ when $\Delta>\frac{d-1}{2}$. Therefore at order one times, the bulk-cone will dominate over the light cone. On the other hand, at late times we expect that the strength of the bulk cone decays to zero, consistent with thermalization of observables localized away from the boundary. \\
\indent Let us now see this decay explicitly. The time dependence of the singularity takes the form 
\begin{align}
G_W(t,\theta)\propto \frac{(u(t)^2-1)^{\Delta-d/2}}{(u(t))^{2\Delta-\frac{d-1}{2}}\sqrt{T'(u(t))}}\frac{1}{(t_{\text{BC}}(2\pi j\pm \theta)-t+i\epsilon)^{2\Delta-\frac{d-1}{2}}}.
\end{align}
At late times, $u(t)$ approaches $\Omega$, so the leading time-dependence comes from $(T'(u(t))^{-\frac{1}{2}}$. For $u\sim \Omega$, we can approximate the formula (\ref{tofu}) for $T(u)$ as follows, 
\begin{align}
T(u)\sim -\frac{1}{\gamma}\log(\Omega-u)+\text{constant},
\end{align}
where the Lyapunov exponent $\gamma$ is given by (\ref{eq:LyapunovDef}). It follows that 
\begin{align}\label{ldecay}
\frac{1}{\sqrt{T'(u(t))}}=\text{constant}\times e^{-\frac{\gamma }{2}t},
\end{align}
which decays exponentially at late times with rate $\gamma$. We plot explicitly $u(t)$ and $T'(u(t))$ as a function of $t \geq \pi$ in Figure \ref{fig:Tandu}. Notice that $T'(u(t))$ quickly approaches $e^{\gamma t}$.

\indent We see from (\ref{ldecay}) that the time scale for the decay of the bulk-cone singularity is governed by the Lyapunov exponent associated with the instability of the photon sphere. This fact is very familiar in astrophysics \cite{ames1968optical} and was also recently discussed in \cite{Polchinski:2015cea,Hadar:2022xag,Kapec:2022dvc}. The basic intuition is that a wave which is initially localized near the photon sphere quickly spreads out away from the photon sphere region due to the instability, and can easily fall into the black hole. Note that a similar decay of the singularity residue of the two-point function with time controlled by the Lyapunov exponent was observed in asymptotically flat black holes as well \cite{Dolan:2011fh}, see formula (42) in that paper, where $\mathcal{X} \sim e^{- T/[2 \sqrt{27}]}$. Using $\gamma = {1 \over \sqrt{27}}$ in the units $M=1$, this indeed coincides with the expected behavior $e^{- \gamma t/2}$.
\\\indent Let us now discuss the late time behavior of the correlator. We have seen that the bulk-cone without bounces is negligible at late times. The strength of the boundary light-cone in $d>2$, see Eq. (\ref{boundarylc}), is constant in time. This is in stark contrast with the case of $d=2$, where it decays exponentially with time, see Appendix \ref{app:BTZ}.\footnote{One can think of this exponential decay as a consequence of the fact that the AdS boundary at $r=\infty$ acts as an unstable photon sphere in the BTZ black hole.} Instead, in $d>2$ we expect that the presence of the black hole horizon narrows the light-cone singularity at late times, see Appendix \ref{app:latetimelightcone}. In addition to the boundary light-cone, the bouncing singularities could potentially contribute as $t\to \infty$. Since we were unable to predict the size of the bounces, we cannot compute their late time behavior. However, numerics suggest that bounce singularities do not decay to zero, so a full understanding of the late-time structure of the correlator would need to take bounces into account. We leave this important problem to future work.

\section{Singularities from numerics}\label{sec:numerics}

In this section we study singularities of Lorentzian two-point retarded correlators numerically by solving the corresponding wave equations in the bulk in $d=3$ and $d=4$. In all examples, we reproduce the expected location of the singularities and successfully match the relative strength of the first winding singularities to the predictions in the previous section.

\subsection{Correlator in momentum space}

In order to numerically compute the retarded correlator on the sphere, we solve the wave equation \eqref{waveeqbrane} with the potential given by \eqref{potential}.  We use the $\mathtt{NDSolve}$ function in $\mathtt{Mathematica}$, imposing ingoing boundary conditions at the horizon (see for example \cite{Kovtun:2006pf,Hartnoll:2009sz}). At the boundary the solution looks like (e.g. $d=4$ and $\Delta = 4$)
\be
\begin{aligned}
&\psi_{\omega \ell}  (v) = \mathcal{B}(\omega,\ell) \psi_\mathcal{B}(v) + \mathcal{A}(\omega,\ell)\psi_\mathcal{A}(v) \,, \\
&\psi_\mathcal{B}(v) = v^2\left(1+b_1 v+ b_2 v^2 + \mathcal{O}(v^3)\right) \,, \\
&\psi_\mathcal{A}(v) = h_a \psi_\mathcal{B}(v) \log v + 1+a_1 v + a_2 v^2 + \mathcal{O}(v^3) ,
\end{aligned}
\ee
in the convenient coordinate $ 1 \geq v \equiv r_+^2/r^2 \geq 0$. Note that $a_2$ is left  as a free parameter since the two solutions can mix from $\mathcal{O}(v^2)$ on. We set $a_2 =0$, and according to \eqref{grdef} $G_R$ is given by
\be\label{eq:GRnumc}
G_R(\omega,\ell) = \lim_{v\to0} \left( \frac{\psi_{\omega \ell}''(v)}{2\psi_{\omega \ell}(v)}- h_a \left(\log v + \frac{3}{2}\right) \right) \,.
\ee
When computing $G_R(\omega,\ell)$ numerically, we introduce a finite cutoff close to the horizon $v=1 - \epsilon_{\rm H}$ where we impose the purely ingoing boundary condition, and we read off the boundary correlator from \eqref{eq:GRnumc} at finite $v=\epsilon_{\rm B}$. Below we set $\epsilon_B = \epsilon_H = 10^{-6}$, and we have checked that our results are stable against changing the cutoffs. We further choose small values of $\mu$ ($\mu=\frac{1}{50}$ in $d=4$ and $\mu=\frac{1}{15}$ in $d=3$) in order to cleanly separate the singularities. However, conceptually nothing is different for $\mu>2$ (above the Hawking-Page transition). 
\begin{figure}[h]
    \centering
    \includegraphics[width=0.75\textwidth]{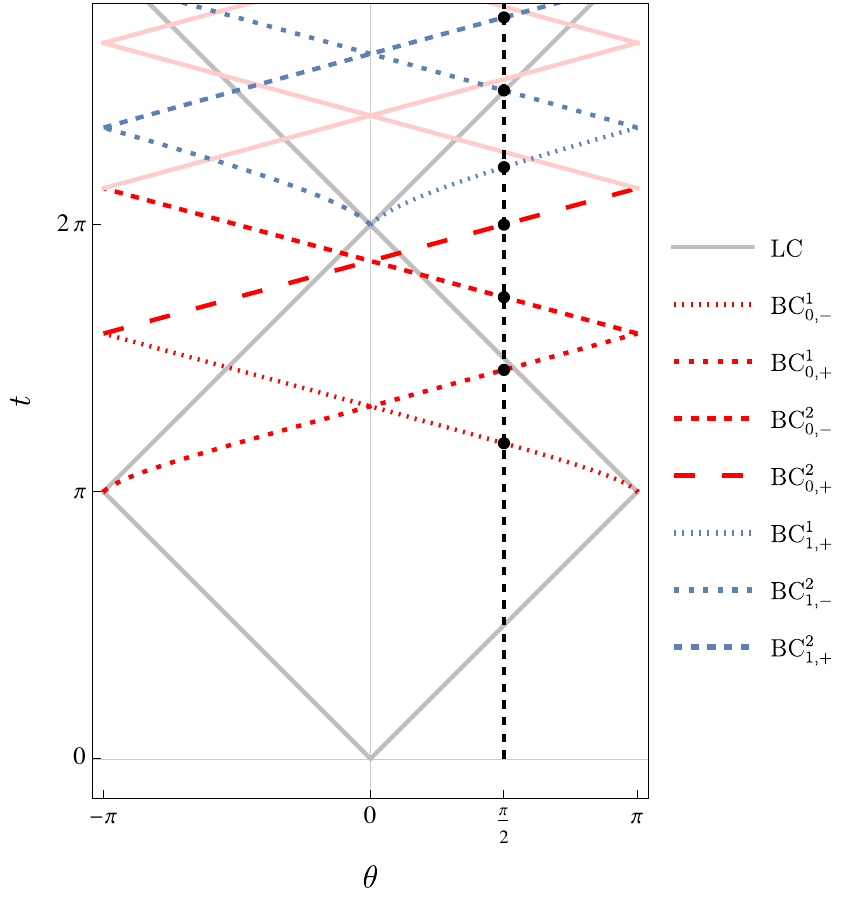}
  \caption{Here we show the analog of Figure \ref{fig:bulkConeInt} for $\mu=\frac{1}{50}$ in $d=4$. The $n$-bounce bulk-cone singularity BC$_{n,\pm}^j$ is defined by \eqref{eq:BounceDef}. The black dashed line corresponds to the slice $\theta=\frac{\pi}{2}$ that was used in our numerical computations and the black dots indicate the predicted bulk-cone singularities that were shown in Figure \ref{fig:bulkConeInt}. The gray lines correspond to the boundary light-cone. The solid weak red lines correspond to further windings of the no-bounce singularity, whose residues decay exponentially with a rate determined by the Lyapunov exponent and are therefore not visible in our numerics compared to the bouncing singularities.  Note that with this value of $\mu$ the slope of the bulk-cone singularities is small compared to 1, and therefore as time increases we see windings of the no-bounce singularity before we hit the first bouncing singularity.}
    \label{fig:spaceTimePrediction}
\end{figure}

\subsection{Correlator in position space}

To numerically compute the Fourier transform, it is convenient to deform the contour to $\text{Im }\omega=\delta>0$ in order to move further away from quasi-stable orbit resonances. This is allowed because $G_R$ is analytic in the upper half $\omega$ plane. Since $G_R(-\omega,\ell) = G_R(\omega,\ell)^*$, we can conveniently write down the $\omega$-integral as
\be\label{eq:foldingint}
\int_{-\infty+i \delta}^{\infty + i \delta} d \omega \, G_R(\omega, \ell) e^{-i\omega t} = \int_{0+i\delta}^{\infty + i \delta} d \omega \, \left(G_R(\omega, \ell) e^{-i\omega t} + G_R(\omega, \ell)^* e^{i\omega^* t} \right) \,.
\ee

To numerically evaluate the integral and the sum over $\ell$ we discretize the contour with a spacing $\delta \omega$, and introduce UV cut-offs $\omega_{\text{max}}$, $\ell_{\text{max}}$ and smoothing factors $e^{-\ell^2/\ell_c^2} e^{-\left(\text{Re }\omega\right)^2/\omega_c^2}$ following \cite{Casals:2009zh, Buss:2017vud}. We set $\omega_{\text{max}} \gg \omega_c \gg 1$ and $\ell_{\text{max}} \gg \ell_c \gg 1$. This converts the light- and bulk-cone singularities into finite bumps and removes spurious oscillations introduced by the UV cut-offs in the transform. Physically, the smoothing factors introduce smearing  of the correlator on the time scale $\delta t \sim {1 \over \omega_c}$ and angular scale $\delta \theta \sim {1 \over \ell_c}$.
\begin{figure}[h]
\centering
\includegraphics[width=\textwidth]{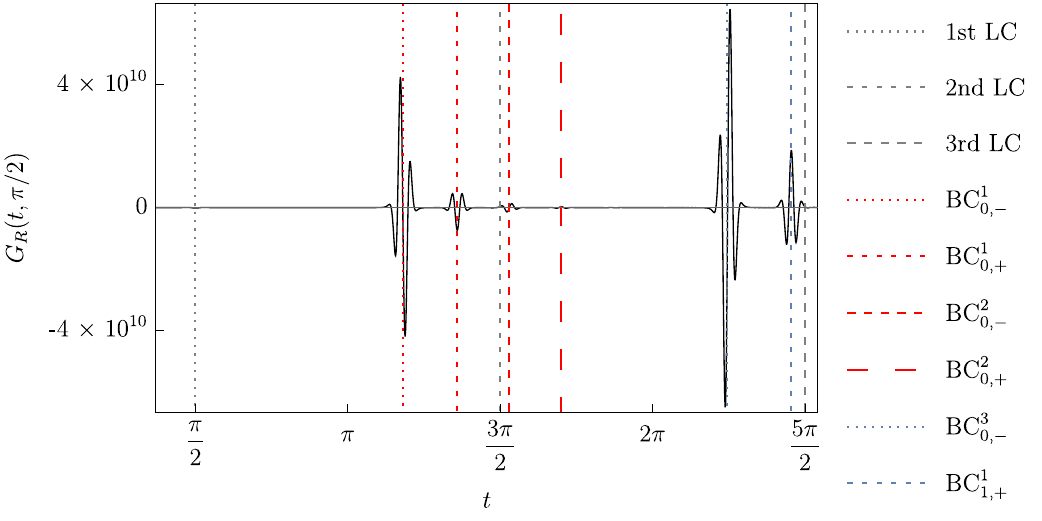}
\caption{$G_R(t,\pi/2)$ in $d=3$ with $\Delta=3$ and $\mu=1/15$. Grey dashed lines correspond to light-cone singularities, which are present but are difficult to see at this scale. The red and blue dashed lines correspond to the analytic predictions from \eqref{eq:BounceDef} for the position of the bulk cone singularities. Further windings are there as well, but are almost invisible at these scales. Here $\omega_c = \ell_c = 40$, $\delta=1/5$, and $\omega_{\text{max}} = \ell_{\text{max}} = 150$.}
\label{fig:GR3d}
\end{figure}

\begin{table}[h]
\centering
\begin{tabular}{ |c|c|c|c| } 
\hline
  \, & Ratios & Numerics & Analytic prediction  \\ 
 \hline
 \, & BC$_{0,+}^1$/BC$_{0,-}^1$ & \textbf{0.1}730 & \textbf{0.1}865 \\ 
 3d & BC$_{0,-}^2$/BC$_{0,-}^1$ & \textbf{0.03}25 & \textbf{0.03}10\\ 
 \, & BC$_{0,+}^2$/BC$_{0,-}^1$ & \textbf{0.0073} & \textbf{0.0073} \\ 
\hline
\, & BC$_{0,+}^1$/BC$_{0,-}^1$ & \textbf{0.09}54 & \textbf{0.09}46 \\ 
 4d & BC$_{0,-}^2$/BC$_{0,-}^1$ & \textbf{0.010}4 & \textbf{0.010}2 \\ 
 \, & BC$_{0,+}^2$/BC$_{0,-}^1$ & \textbf{0.0011} & \textbf{0.0011} \\ 
 \hline
\end{tabular}
\caption{Comparison of winding bulk-cone peaks between numerics and the analytic prediction from \eqref{eq:GWleadBC} in 3d and 4d. All peaks are normalized by the first bulk-cone.}
\label{tab:3dpeaks}
\end{table}
\subsection{Results}
\begin{figure}[t]
     \centering
     \begin{subfigure}[b]{0.48\textwidth}
         \centering
    \includegraphics[width=\textwidth]{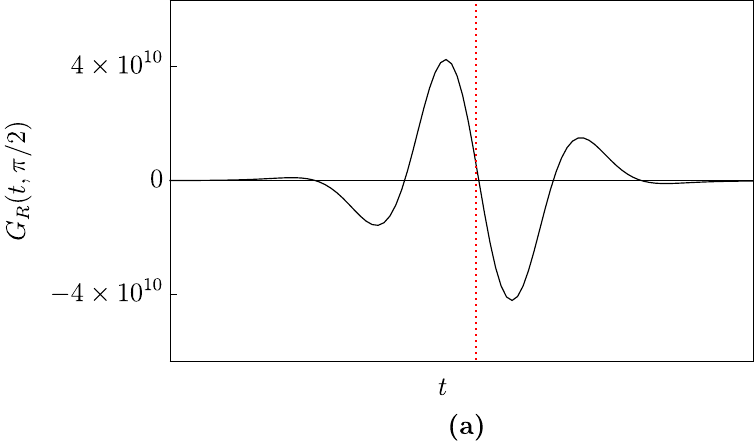}
     \label{fig:3dbc1}
     \end{subfigure}
     \hfill
     \begin{subfigure}[b]{0.48\textwidth}
         \centering
    \includegraphics[width=\textwidth]{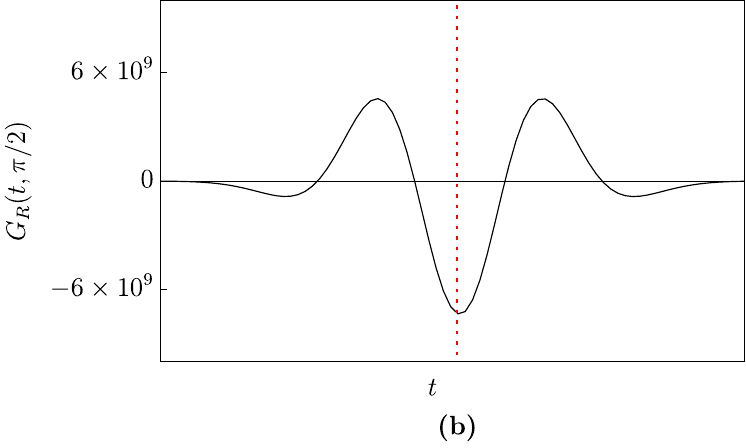}
     \label{fig:3dbc2}
     \end{subfigure}
     \hfill
     \vspace{0.5cm}
     \begin{subfigure}[b]{0.48\textwidth}
         \centering
    \includegraphics[width=\textwidth]{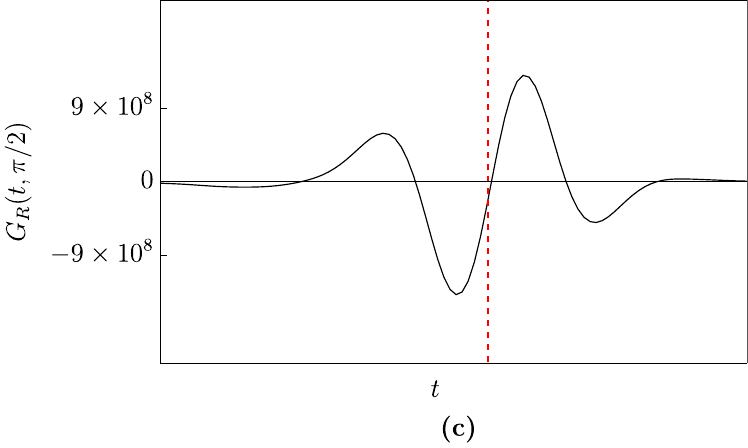}
    \label{fig:3dbc3}
     \end{subfigure}
     \hfill
     \begin{subfigure}[b]{0.48\textwidth}
         \centering
    \includegraphics[width=\textwidth]{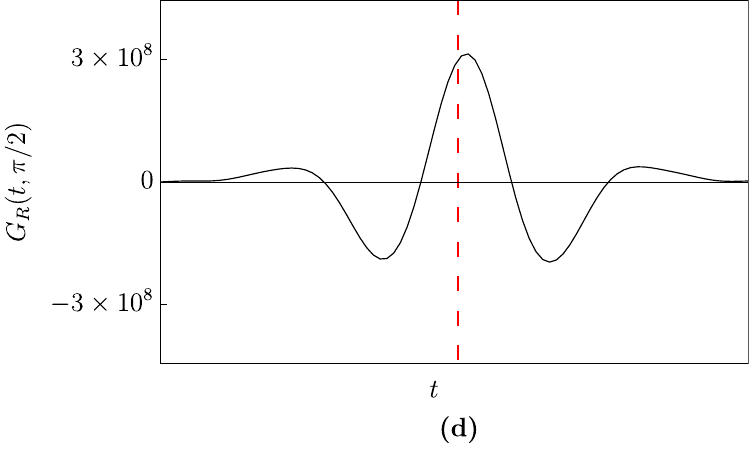}
     \label{fig:3dbc4}
     \end{subfigure}
\caption{$G_R(t,\pi/2)$ in 3d close to (a) BC$_{0,-}^1$, (b) BC$_{0,+}^1$, (c) BC$_{0,-}^2$, (d) BC$_{0,+}^2$. The bulk-cone bumps exhibit the four-fold structure discussed in Section \ref{sec:dimdep}. Since $G_R = -2\theta(t) \text{Im }G_W$, the relative $i$ between BC$_{0,-}^k$ and BC$_{0,+}^k$ coming from $e^{i\pi\frac{d-2}{2}}$ in \eqref{eq:GWleadBC} changes the shape of the bumps. The shapes of BC$_{0,-}^k$ are consistent with the prediction $\sim \text{Im}\,i[t_{\text{BC}}(2\pi k-\frac{\pi}{2})-t+i\epsilon]^{-5}$ and those of BC$_{0,+}^k$ with $\sim \text{Im}\, [t_{\text{BC}}(2\pi k+\frac{\pi}{2})-t+i\epsilon]^{-5}$. The shapes of BC$_{0,\pm}^1$ and BC$_{0,\pm}^2$ are the same up to a minus sign. The parameters of these plots and legends for the dashed lines are the same as in Figure \ref{fig:GR3d}.}
\label{fig:4fold}
\end{figure}
\begin{figure}
\centering
\includegraphics[width=\textwidth]{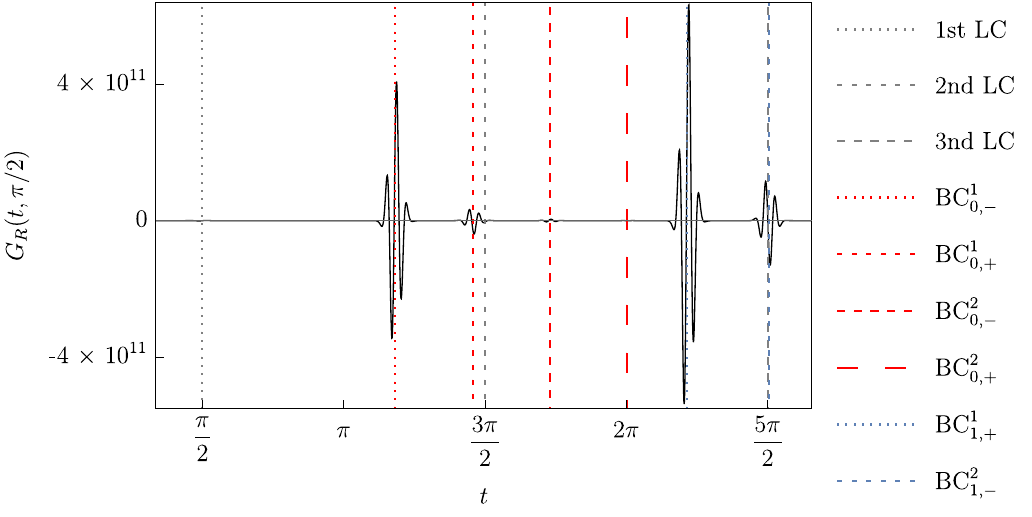}
\caption{$G_R(t,\pi/2)$ in $d=4$ with $\Delta=4$ and $\mu=\frac{1}{50}$. The grey dashed lines are the light-cone singularities, which are present but are difficult to see at this scale in comparison to the bulk-cone singularities. The red dashed lines correspond to the analytic predictions from \eqref{eq:BounceDef} for the position of the first four bulk cone singularities from the BC$_0$ series. Further windings will overlap with bouncing singularities and therefore will be difficult to distinguish., The dashed blue lines correspond to the predictions from the first bouncing singularity and its first few windings. Here $\omega_c = \ell_c = 35$, $\delta=1/5$, and $\omega_{\text{max}} = \ell_{\text{max}} = 150$. Note that the residues at the windings BC$_{0,\mp}^{2}$ are suppressed by $10^{-2}$ and $10^{-3}$, respectively, compared with the residue at BC$_{0,-}^1$ and are therefore difficult to see at this scale.}
\label{fig:GR4d}
\end{figure}

We have implemented the scheme above in $d=3$ and $d=4$. We plot the outcome of the computations in Figures \ref{fig:GR3d} and \ref{fig:GR4d} respectively. Here we denote the $(n-1)$ bounce bulk-cone singularity (\ref{eq:BounceDef}) by BC$_{n-1,\pm}^j$. Given fixed $(\theta,n,j,\pm)$, we can solve $2\pi j\pm \theta=n\Theta(u)$ for $u$, and then insert the solution into $t=nT(u)$ to obtain the time $t$ of the corresponding singularity. The prediction for the locations of the singularities on the boundary in $d=4$ is shown in Figure \ref{fig:spaceTimePrediction}, and the $d=3$ case is qualitatively similar.

Let us start with the correlator in $d=3$. We set $\Delta=3$, $\mu =1/15$, $\theta =\pi/2$, and we study the correlator as a function of $t$. We find the following results:
\begin{itemize}
\item The positions of light-cone and winding bulk-cone singularities are consistent with the expected locations. For the light-cone singularities these are $t=\theta+2\pi m$ and $t=2\pi-\theta+2\pi m$ with $m=0,1,\ldots$. For the bulk-cone singularities these are given by \eqref{eq:BounceDef}.
\item As shown in Figure \ref{fig:4fold}, the bulk-cone singularities exhibit the four-fold structure predicted from \eqref{eq:GWleadBC} with $d=3$, see also Section \ref{sec:dimdep}.
\item Bulk-cone bumps are higher then light-cone ones, consistent with the fact that $G_R$ diverges as $\delta t^{-(2 \Delta -{d-1 \over 2})} = \delta t^{-5}$ at bulk-cone points and as $\delta t^{-\Delta} = \delta t^{-3}$ on the light-cone.
\item The ratios of heights of winding bulk-cone peaks are always within $\sim 10 \%$ of the prediction in \eqref{eq:GWleadBC}. We show the comparison in Table \ref{tab:3dpeaks}.
\item The bumps at the dashed blue lines in Figure \ref{fig:GR3d} correspond to bulk-cone singularities that include bounces from the AdS boundary. The location of these bouncing singularities is consistent with (\ref{eq:BounceDef}) with $n=2$, but our attempt to use \eqref{eq:GRimspbh} to reproduce the shape and size of the bounces did not produce the expected results. Note that the height of the first bounce is actually larger than any of the no-bounce winding singularities.
\end{itemize}

\begin{figure}
     \centering
     \begin{subfigure}[b]{0.49\textwidth}
         \centering
    \includegraphics[width=\textwidth]{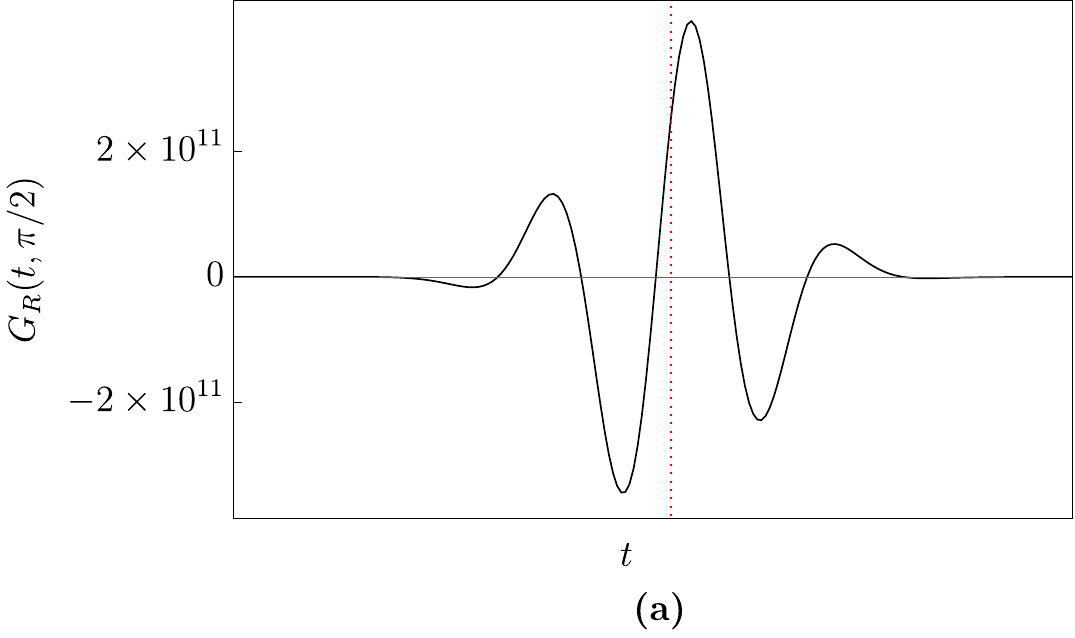}
    \label{fig:4dbc1}
     \end{subfigure}
     \hfill
     \begin{subfigure}[b]{0.49\textwidth}
         \centering
    \includegraphics[width=\textwidth]{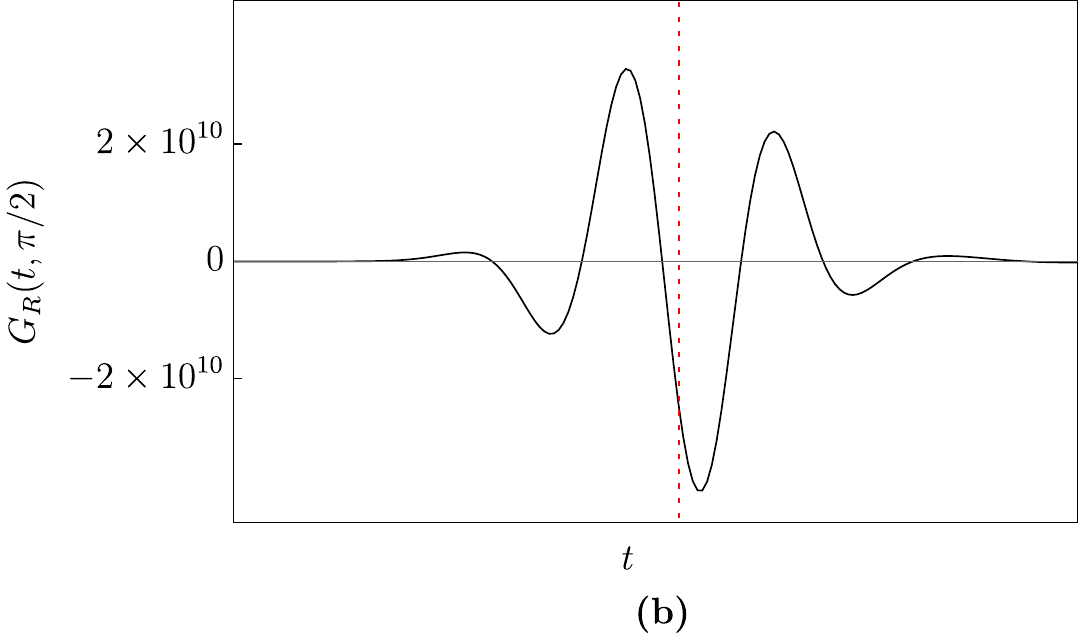}
    \label{fig:4dbc2}
     \end{subfigure}
\caption{$G_R(t,\pi/2)$ in 4d close to (a) BC$_{0,-}^1$, (b) BC$_{0,+}^1$. The bulk-cone bumps exhibit the two-fold structure discussed in Section \ref{sec:dimdep}. Since in 4d $e^{i\pi\frac{d-2}{2}} = -1$ in \eqref{eq:GWleadBC}, shapes of BC$_{0,-}^1$ and BC$_{0,+}^1$ are the same up to a minus sign. The parameters of these plots and legends for the dashed lines are the same as in Figure \ref{fig:GR4d}. The shapes and locations are consistent with the prediction $\sim \text{Im}\,[t_{\text{BC}}(2\pi\pm\frac{\pi}{2})-t+i\epsilon]^{-\frac{13}{2}}$.}
\label{fig:2fold}
\end{figure}

Next we repeat the same exercise in $d=4$. We set $\Delta=4$, $\mu =\frac{1}{50}$, and $\theta =\frac{\pi}{2}$ and we study the correlator as a function of $t$. The results are as follows:
\begin{itemize}
    \item We find light-cone and bulk-cone singularities at the predicted locations. 
    \item As shown in Figure \ref{fig:2fold}, the bulk-cone singularities exhibit the two-fold structure predicted in $d=4$.
    \item $G_R$ diverges as $\delta t^{-(2 \Delta -{d-1 \over 2})} = \delta t^{-\frac{13}{2}}$ on the bulk-cone and only as $\delta t^{-\Delta}=\delta  t^{-4}$ on the light-cone. Accordingly bulk-cone bumps are higher than light-cone ones.
    \item The ratios of heights of bulk cones are in good agreement with the prediction in \eqref{eq:GWleadBC}. The comparison is shown in Table \ref{tab:3dpeaks}.
    \item The position of bouncing singularities is correctly predicted by including $e^{-2n S(0,z_-)}$ corrections to the residues. However, the shape and size of these singularities is not correctly reproduced by \eqref{eq:GRimspbh}.
\end{itemize}

\section{Stringy and gravitational corrections}
\label{sec:string}

Our computation so far was done in the approximation where the boundary operator is described by free wave propagation on the fixed AdS Schwarzschild background. In the language of the CFT dual this is a good description in the planar limit when the 't Hooft coupling is arbitrarily large. As we move away from this holographic limit new physical effects arise.\\
\indent Let us first discuss stringy effects, which should become important as we decrease the 't Hooft coupling away from infinity (in other words, we consider finite $\lambda$ effects). First, there are stringy corrections to the black hole geometry itself. However, as long as $r_s \gg \sqrt{\alpha'}$ we expect that these are small. Second, there are effects related to string propagation on the Schwarzschild background. These were analyzed in \cite{Dodelson:2020lal}. More precisely, to understand stringy corrections to the singularity, we need to study string propagation along null geodesics in the bulk. This is given by the so-called pp-wave limit \cite{Blau}, which is solvable. The basic physical effect is tidal excitation of the string, which attenuates the singularity and turns it into a finite bump of width $\Gamma(\alpha',t)$.\footnote{This attenuation is clearest for the light-cone singularity in the bulk-to-bulk propagator, but there are subtleties involved in computing the boundary two-point function, see the discussion in \cite{Dodelson:2020lal}. Here we assume that these subtleties are unimportant.} \\
\indent 
At early times $t\sim \pi$ the effective width takes the form
\be
\Gamma(\alpha',t) \propto 
{\alpha' G M \over \Big(r_+(u(t)) \Big)^{d}},
\ee
where $r_+$ is the outermost turning point. The relevant physics describes propagation through a gravitational shock wave, so that the computation mimics the one of \cite{Amati:1987uf}, with a familiar formula for tidal excitations $\text{Im } \delta_{\text{tidal}} \sim {\alpha' G s  \over b^{D-2}}$.\footnote{Here $D=d+1$ is the dimensionality of the bulk.} For the same reason we expect this tidal resolution of the singularity to be universal for any extended object that has internal excitations, e.g. a hydrogen atom.\footnote{It would be interesting to do the computation for the hydrogen atom explicitly, see \cite{Parker1980,Parker1982,Gill1987} for related work.}

At late times the effective width takes the form 
\be
\Gamma(\alpha',t) \propto  {\alpha' \over (G M)^{\frac{2}{d-2}}} \log {1 \over {r_+(u(t)) \over r_-(u(t))}-1} ,
\ee
where $r_+$ and $r_-$ are the turning points just outside and just inside the photon sphere respectively. Here the tidal force becomes so strong that it effectively rips the object apart (by extending it along one transverse direction and compressing along others). This again we expect to be quite universal and the effect kicks in when the tidal force becomes comparable to the binding force that keeps it together. Tidal excitations in thermal CFTs were also recently explored in \cite{Engelsoy:2021fbk}.

Let us next briefly discuss gravitational effects (or finite $N$ corrections). The simplest one to consider is emission of gravity waves.
Indeed, as a particle spirals around the black hole it will emit gravity waves, and therefore we expect the exclusive amplitude which our two-point correlator computes to be suppressed. Considering for simplicity propagation through a gravitational shockwave in flat space, the relevant correction to the phase shift takes the form $\text{Im } \delta_{\text{GW}} \sim {G^3 s^2  \over b^{3D-10}}$ \cite{Amati:1987uf}. It is an interesting question whether gravitational effects eventually completely remove the singularity from the complex plane or not.

To conclude we expect the bulk-cone singularities to be absent at finite $N$ and $\lambda$, but the corresponding features, namely the bulk-cone bumps, should remain. In fact, based on this discussion, it seems very natural that at finite coupling the only true singularities of the thermal two-point function on the sphere are light-cone singularities. 

\section{Photon rings versus black holes}
\label{sec:prvsbh}

The fact that astrophysical black holes are surrounded by a photon shell is one of their key properties, which is responsible for many of their observable signatures as reviewed in Appendix~\ref{sec:Signatures}.
The existence of the photon shell crucially relies on the compactness of the hole, and it is widely believed that \textit{only} black holes are sufficiently compact to lie within their photon shell, and hence capable of producing a photon ring in their image.
We now briefly review some of the evidence for this claim. 

Here we consider four-dimensional black holes in asymptotically flat spacetime. Recall that a Schwarzschild black hole has an event horizon radius of $r=2M$, which is well within its photon sphere located at $r=3M$.

In 1959, Buchdahl \cite{Buchdahl1959} proved that, under certain mild assumptions, a static and spherically symmetric matter configuration of total mass $M$ must occupy a region of space with radius $R>(9/4)M$, thus making precise the idea that ordinary matter cannot reach an arbitrarily high density before collapsing into a black hole.
Although Buchdahl's theorem does not rule out the possibility that an ultradense object (such as a neutron star) could be sufficiently compact to lie within its photon sphere, Buchdahl's bound has since been greatly improved via different methods.

Building upon the seminal work of Hartle \cite{Hartle1978}, a much stronger and rather general bound was eventually obtained in 1984 by Lindblom \cite{Lindblom1984}, who used causality constraints to argue for a compactness limit $R\gtrsim 2.8M$ on neutron stars.
More precisely, Hartle, Lindblom, and others assumed the equation of state to be known (from nuclear theory and experiments) up to some nuclear density $\rho_0$, and then used the requirement that the speed of sound within the star be subluminal (the causal bound $dp/d\rho\leq c^2$) to constrain the allowed total mass and size of the star, and hence to bound its compactness.
This calculation is weakly sensitive to the specific choice of $\rho_0$ and the precise equation of state assumed for $\rho\leq\rho_0$ (which in practice is not quite perfectly known), but the causal limit $R\gtrsim 2.8M$ remains quite robust.
Though it does technically leave some region of parameter space for a neutron star to have a photon shell, the state of the art suggests that all such models are very contrived.
Indeed, all of the modern, realistic equation-of-state calculations rule out this possibility, as can be clearly seen in Figure~7 (right panel) of the recent review \cite{Ozel:2016oaf}.
This is ultimately the strongest evidence ruling out neutron stars with a photon ring.

Once neutron stars are ruled out, the only possible remaining loopholes to the dictum that ``only black holes have a photon ring'' are exotic ultracompact objects (such as boson stars) whose existence is highly speculative.
Even in that context, it seems very difficult to engineer configurations dense enough to have a photon ring (see, e.g., \cite{Cardoso:2021ehg}).
More sophisticated arguments have been developed to rule out such configurations as well, even lifting the restriction of spherical symmetry (but still assuming stationarity and axisymmetry).
For instance, a particularly promising line of attack proceeds from the observation that ``photon spheres always come in pairs'' for ultracompact objects that are not black holes, and that moreover, an unstable photon sphere (which is needed for light to escape and produce a photon ring) is always accompanied by another stable photon sphere \cite{Cunha:2017qtt}.
This result is significant because the instability of the bound photon orbits in the Kerr spacetime is a necessary condition for the stability of the Kerr family of metrics under small perturbations.
Conversely, stable photon spheres have been argued to generally lead to nonlinear spacetime instabilities, as they can keep accruing massless particles (or trapping waves) until enough energy density has accumulated to backreact on the geometry and collapse the compact object into a black hole.
Despite some initial doubts about the onset of such instabilities \cite{Sanchis-Gual:2019ljs}, more recent numerical investigations seem to suggest that ultracompact objects with an outer photon shell are indeed unstable, either to black hole collapse or expansion to non-compact configurations without a photon ring \cite{Cunha:2022gde}.
This line of argument may thus rule out even exotic configurations with a photon shell.\footnote{It would be interesting to study the existence and stability of such configurations within AdS/CFT.}

At any rate, discarding these exotic possibilities, there is substantial evidence that the only compact astrophysical objects with unstably bound photon orbits---and hence a photon ring---are black holes.
If this belief indeed holds true, then it is of great empirical importance and can help tackle the question: ``How can one ascertain whether an astrophysical source is truly a black hole?''

The photon ring provides an operational answer: it is present if (and likely, only if) the source is a black hole.
In other words, measuring a photon ring around an astrophysical object could not only provide a consistency test, but also a smoking gun signature, for the Kerr nature of the source.
In the context of this paper, it would be very interesting to understand to what extent black holes are the only objects that possess a photon sphere within AdS/CFT. To phrase this question in the language of CFT: what is the set of states in the boundary theory exhibiting a bulk-cone singularity?

\section{Conclusions}\label{sec:conclusions}

In this paper we analyzed singularities of the holographic real-time thermal two-point function $\la {\cal O}(t,\theta) {\cal O}(0) \ra_{S^1 \times S^{d-1}}$ on the spatial sphere. This correlator is dual to wave propagation in an AdS Schwarzschild black hole background. In addition to the usual light-cone singularities it exhibits an interesting pattern of bulk-cone singularities captured by null geodesics in the black hole geometry.
 
We analytically derived the leading bulk-cone singularity for the two-point function of scalar operators. The result is pictorially summarized by Figure \ref{fig:bulkConeInt}, see \eqref{eq:GWleadBC} for the precise formula. A striking feature of the black hole bulk-cone singularities is that they exhibit group velocity larger than the speed of light. They originate from geodesics wrapping the photon sphere in the bulk.  Our result \eqref{eq:GWleadBC} exhibits several interesting features: 
\begin{itemize}
    \item For $\Delta>{d-1 \over 2}$ it is more singular than the ordinary light-cone singularity.
    \item The coefficient of the singularity decays with time, with rate controlled by the Lyapunov exponent of null geodesics around the photon sphere.
    \item The bulk-cone singularities exhibit an $N(d)$-fold structure, with $N(3)=4$ and $N(4)=2$, see Section \ref{sec:dimdep}. 
\end{itemize}
This result extends the analysis of \cite{Hubeny:2006yu,Dodelson:2020lal} beyond the geodesic approximation. It also represents the AdS analog of the flat space results \cite{Dolan:2011fh,Buss:2017vud,Casals:2016qyj,Zenginoglu:2012xe}.

We also computed the correlator numerically. In this case to get reliable results we effectively smeared the correlator in space and time, which turns singularities into finite-size bumps. We tested our analytic predictions numerically and confirmed them with the available precision. We also observed singularities due to bouncing geodesics, which were predicted in \cite{Dodelson:2020lal}. 

Stringy and gravitational corrections are expected to remove the bulk-cone singularities and turn them into bumps. It would be interesting to understand if these stringy features can be reproduced using the thermal product formula \cite{Dodelson:2023vrw}, or equivalently, stringy quasi-normal modes, see e.g. \cite{Casalderrey-Solana:2018rle,Grozdanov:2018gfx}. These black hole bulk-cone bumps provide a clear boundary signature of the photon sphere at large but finite $N$ and $\lambda$. In the context of the physics of gravitational waves, the retarded two-point function enters the computation of the self-force which affects the worldline of the inspiraling compact body, see e.g. \cite{Casals:2009zh,Casals:2013mpa,Wardell:2014kea}. It would be interesting to explore how the change in the singularity structure of the two-point function affects the waveform of emitted gravity waves.

There are several interesting directions in which our analysis could be extended and improved. While in our numerical analysis we have observed bulk-cone singularities that include bounces from the AdS boundary we have not derived analytically the form of the leading singularity in this case. Notice that from our numerical analysis it is clear that bounce singularities become dominant at later times. It would be very interesting to understand their late-time structure in more detail. 
A related question is how to go beyond the leading singularity computed in this paper, which would require several technical improvements of our analysis.

In our analytic investigation we found it useful to organize the computation in terms of thermal Regge poles (which are poles in spin of the correlator at a given Matsubara frequency). Regge poles with the smallest value of $\text{Im } \ell$ dominate the late-time behavior of the no-bounce bulk-cone singularity. It would be interesting to understand the structure of thermal Regge poles in a generic CFT, and if they have a clear physical interpretation.

It would be interesting to understand if techniques similar to our numerical analysis of the smeared correlator can be used to compute real-time correlators in CFTs using the spectrum of low-lying operators and their three-point functions. Indeed, by inserting a complete set of states we can express $\la {\cal O}(t,\theta) {\cal O}(0) \ra_{S^1 \times S^{d-1}}$ as a sum over three-point functions, see e.g.\ \cite{Festuccia:2006sa,El-Showk:2011yvt}. Similarly, it would be also interesting to see if the bootstrap \cite{Iliesiu:2018fao,Iliesiu:2018zlz},  lattice \cite{Brower:2012vg,Brower:2020jqj}, or fuzzy sphere approach \cite{Zhu:2022gjc,Hu:2023xak,Han:2023yyb} can be used to study the thermal correlator on the sphere at real times.

An obvious extension of our work is to consider more general backgrounds. For example we can consider non-zero angular velocities \cite{Hawking:1998kw}.  In this case (within a certain parameter range \cite{Kim:2023sig,Benjamin:2023qsc}) the dual geometry becomes AdS Kerr and the structure of singularities depends on the orientation on the sphere. For instance, right-moving and left-moving singularities in the equatorial plane will exhibit different late-time velocities.
It would be interesting to determine whether the recently identified emergent conformal symmetry of the Kerr photon shell \cite{Hadar:2022xag}, which must also arise in AdS Kerr, has a clearer holographic interpretation in the context of AdS/CFT.
We can also consider non-trivial chemical potential for charge, and correspondingly explore Lorentzian singularities in large charge EFTs \cite{Hellerman:2015nra,Monin:2016jmo,Loukas:2018zjh}. 

Our analysis was done for $S^1 \times S^{d-1}$, but similar effects are expected to be present on more general backgrounds, as well as away from equilibrium. One well-known example of this type is provided by the so-called Robinson-Trautman spacetimes \cite{BernardideFreitas:2014eoi,Skenderis:2017dnh}. Another very interesting setup was considered recently in \cite{Horowitz:2023ury}, which in the context of the present paper would correspond to studying the structure of the bulk-cone singularities in the presence of shock waves. In \cite{Pantelidou:2023pjn} an excited thermal state was constructed using a Euclidean path integral with a relevant deformation. 

Note that to observe the effects considered in this paper it was important to consider $d>2$. We are not aware of examples of similar effects in lower-dimensional systems.\footnote{Curiously, an analog of the photon sphere has been discussed also for 2+1 dimensional acoustic black holes \cite{Wang:2019zqw}.} It would of course be very interesting to compute thermal correlators on the sphere in higher-dimensional CFTs directly, and to search for signatures of bulk-cone singularities there. In \cite{Amado:2016pgy,Amado:2017kgr} this was done for the singlet sector of free large $N$ gauge theories coupled to vector or adjoint scalar matter, and no bulk-cone singularities were observed. This is consistent with the expectation that the holographic dual in this case is highly nonlocal.

We lack microscopic understanding of the black hole bulk-cone singularities. For example, in \cite{Hatta:2007cs,Hatta:2008tx} a `partonic' picture for the behavior of the two-point function at strong coupling at finite temperature and infinite volume was put forward. Many qualitative features of correlators were captured by assuming that tentative partons (quarks and gluons at strong coupling) that capture gauge theory dynamics are copiously produced democratically distributed across angles. Similarly, it was possible to accommodate the effects of plasma by postulating a certain force that it exerts on partons (its microscopic origin is not understood). Here we see that on the sphere even more peculiar effects appear. In this case some excitations with large enough angular velocity take parameterically large time to thermalize (these correspond to stable orbits \cite{Festuccia:2008zx,Berenstein:2020vlp,Dodelson:2022eiz}). There are other excitations that exhibit anomalous dispersion and group velocity larger than one (these correspond to the bulk cone close to the photon sphere). Finding other examples of this phenomenon in quantum many-body systems would be very interesting.

\section*{Acknowledgments}
We thank Ant\'onio Antunes, David Berenstein, Stefano Giusto, Alba Grassi, Nima Lashkari, Raghu Mahajan, Vasiliy Makhalov, Andrei Parnachev, Hirosi Ooguri, Kyriakos Papadodimas, Elli Pomoni, Marcos Riojas, Rodolfo Russo, Eva Silverstein, Hao-Yu Sun, Aron Wall, and Zahra Zahraee for useful discussions. This project has received funding from the European Research Council (ERC) under the European Union’s Horizon 2020 research and innovation programme (grant agreement number 949077).  The work of CI is partially supported by the Swiss National Science Foundation Grant No. 185723. The work of AL is partly supported by the National Science Foundation Grant No. 2307888.

\appendix

\section{Bounds on the positions of Regge poles}\label{reggebounds}
In this appendix we explore constraints on the position of the Regge poles coming from the wave equation. Like quasi-normal modes, these are defined by the ingoing boundary condition at the horizon and normalizability at the boundary. This discussion follows closely similar arguments for quasi-normal modes \cite{Horowitz:1999jd}. It is convenient to introduce Eddington-Finkelstein coordinates $v=t-z$ in which the metric is given by 
\be 
ds^2=-f(r)\, dv^2+2dv\, dr+r^2\, d\Omega_{d-1}^2.
\ee 
After the Fourier decomposition (\ref{fourierspin}), the wave equation $(\Box-m^2)\phi=0$ is given by 
\be\label{eq:EOMEF}
    f(r)\psi''(r)+(f'(r)-2i\omega)\psi(r)-V_{\text{EF}}(r)\psi(r)=0,
\ee
where 
\be 
V_{\text{EF}}(r) =V_1(r)+V_2(r),
\ee
with 
\begin{align}
V_1(r)&=\frac{(d-1)(d-3)}{4r^2}f(r)+\frac{(d-1)}{2r}f'(r)+\Delta(\Delta-d)\\
V_2(r)&=\frac{\ell(\ell+2\alpha)}{r^2}.
\end{align}
The potential is positive outside the horizon for $d\geq 3$ and $\Delta\notin (\frac{d-1}{2},\frac{d+1}{2})$.
\begin{figure}
    \centering
    \includegraphics[width=0.5\textwidth]{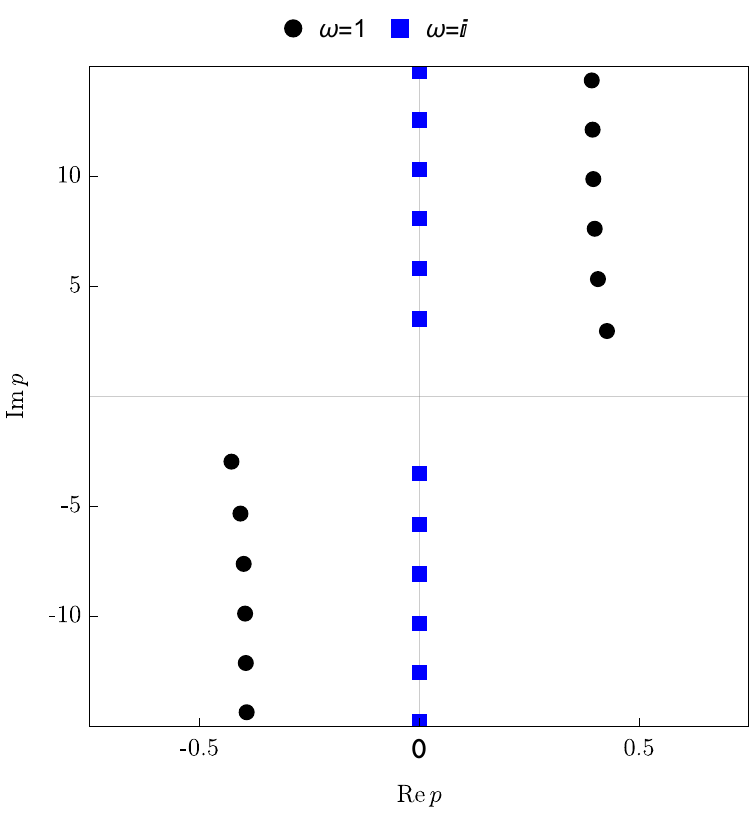}
  \caption{Regge poles for a scalar field in a black hole background obtained numerically from the Mathematica package $\mathtt{QNMSpectral}$ \cite{Jansen:2017oag}. Here $d=4$, $\mu=1$, and $\Delta = 4$. When $\omega$ is real, the Regge poles sit in the first and third quadrant as predicted from \eqref{eq:reggeconst}. When $\omega$ is purely imaginary, the Regge poles are purely imaginary.}
    \label{fig:reggepoles}
\end{figure}

We now multiply the wave equation \eqref{eq:EOMEF} by $\psi^*$ and integrate from the horizon to the boundary to obtain
\begin{align}\label{eq:eomEFBounds}
\int_{r_+}^{\infty}\, dr\left(f|\psi'|^2+2i\omega \psi^*\psi'+(V_1+V_2)|\psi|^2\right)=0,
\end{align}
where we have integrated by parts and used that $f(r_+)=0$ and $\psi^*(\infty)=0$. 

Defining $\ell=\alpha+p$ and taking the imaginary part we get
\begin{align}
(\omega-\omega^*)\int_{r_+}^{\infty}dr\, \psi^*\psi'=\omega^*|\psi|^2(r_+)-\text{Im}(p^2)\int_{r_+}^{\infty}dr \frac{|\psi|^2}{r^2}.
\end{align}
When $\omega$ is real and positive we find
\begin{align}
\text{Im}(p^2)=\frac{\omega|\psi^2|(r_+)}{\int_{r_+}^{\infty}dr\, \frac{|\psi|^2}{r^2}}>0.
\label{eq:reggeconst}
\end{align}
Therefore $p$ must be in either the first or third quadrant, as shown in Figure \ref{fig:reggepoles}. Similarly when $\omega$ is real and negative, $p$ is in the second or fourth quadrant.\\
\indent Secondly, for $\text{Im}\,\omega>0$ let us multiply \eqref{eq:eomEFBounds} by $\omega^*$ and again take the imaginary part, from which we obtain
\begin{align}
\text{Im}(\omega^*(p^2-\alpha^2))&=\frac{\text{Im }\omega\int_{r_+}^{\infty}dr\, f|\psi'|^2+\text{Im }\omega\int_{r_+}^{\infty}dr\, V_1|\psi|^2+|\omega|^2|\psi|^2(r_+)}{\int_{r_+}^{\infty}dr\, \frac{|\psi|^2}{r^2}}>0.
\end{align}
The allowed sectors are then 
\begin{align}
\frac{\text{Arg }\omega}{2} &<\text{Arg}(\sqrt{p^2-\alpha^2})<\frac{\pi+\text{Arg }\omega}{2},\\
\frac{-2\pi+\text{Arg }\omega}{2} &<\text{Arg}(\sqrt{p^2-\alpha^2})<\frac{-\pi+\text{Arg }\omega}{2}.
\end{align}
In fact, for imaginary $\omega$ a numerical computation shows that the Regge poles are purely imaginary, see Figure \ref{fig:reggepoles}.

\section{BTZ}\label{app:BTZ}
In this appendix we study BTZ both at infinite and finite volume as a pedagogical example of how the expansion in terms of Regge poles correctly reproduces the position space correlator and its singularities. We start with the Fourier expansion of the Euclidean correlator at infinite volume ($\beta=2\pi$)
\be
G_E(\tau,x) = \sum_{n=-\infty}^{\infty} e^{in\tau}\int_{-\infty}^{\infty}dk\, e^{ikx}G_{R}(\omega = i|n|, k),
\ee
with the retarded correlator in BTZ given by
\be
G_R(\omega,k) = \frac{\Gamma \left(\frac{\Delta}{2}+\frac{i (k-\omega )}{2}\right) \Gamma \left(\frac{\Delta}{2}-\frac{i
   (k+\omega )}{2}\right)}{4\sin(\pi\Delta)\Gamma (\Delta )^2 \Gamma \left(-\frac{\Delta }{2}+\frac{i (k-\omega )}{2}+1\right) \Gamma \left(-\frac{\Delta}{2}-\frac{i (k+\omega )}{2}+1\right)}.
\ee
Deforming the $k$ contour in the upper half plane, we pick up Regge poles at $k=i(2m+|n|+\Delta)$ and obtain the following representation of the Euclidean correlator,
\be 
G_E(\tau,x) = \sum_{n=-\infty}^\infty\sum_{m=0}^\infty e^{i n \tau-x (\Delta +2 m+|n|)}\frac{\Gamma(m+\Delta)\Gamma (m+\Delta +| n| )}{\Gamma (\Delta )^2\Gamma(m+1)\Gamma (m+| n| +1)}.
\ee
Performing the sums we find the expected expression
\be
G_E(\tau,x) = \frac{1}{2^\Delta(\cosh x-\cos\tau)^\Delta}.
\ee

\indent Consider now instead the Euclidean correlator at finite volume
\be
G_E(\tau,\theta) = \sum_{n,\,\ell=-\infty}^{\infty} e^{in\tau+i\ell\theta}G_{R}(\omega = i|n|, \ell) \ . 
\ee
We rewrite the sum over $\ell$ as an integral using the contours $\mathcal{C}_+$ and $\mathcal{C}_-$ as in Figure \ref{fig:ReggePlane}. Then
 \begin{align}
    G_E(\tau,\theta)=\frac{i}{2}\sum_{n=-\infty}^{\infty} e^{in\tau}\,\oint_{\mathcal{C}_++\mathcal{C}_-}\frac{d\ell}{\sin(\pi \ell)}e^{i\ell(\theta-\pi)}G_R(i|n|,\ell).
 \end{align}
 Using the symmetry under $\ell\to-\ell$ we can write this as an integral to the right above the real axis, 
\begin{align}
  G_E(\tau,\theta)=i\sum_{n=-\infty}^{\infty} e^{in\tau}\int_{\mathcal{C}_+}\frac{d\ell}{\sin(\pi \ell)} \cos(\ell(\pi-\theta))G_R(i|n|,\ell).
\end{align}
We can proceed analogously to the infinite volume case by deforming in the upper half plane  and picking up the Regge poles at $\ell=i(2m+|n|+\Delta)$,
\be \label{eq:BTZfinite}
 G_E(\tau,\theta)=\sum_{n,\,s=-\infty}^{\infty}\sum_{m=0}^\infty e^{i n \tau-|2\pi s+\theta|(\Delta +2 m+|n|)}\frac{\Gamma (m+\Delta ) \Gamma (m+\Delta +| n| )}{\Gamma (\Delta )^2 \Gamma(m+1)\Gamma(m+|n|+1)},
\ee 
where the sum over $s$ comes from the kernel
\begin{align}
    \frac{\cosh((\Delta+2m+|n|)(\pi-\theta))}{\sinh(\pi (\Delta+2m+|n|))}=\sum_{s=-\infty}^{\infty} e^{-|\theta+2\pi s|(\Delta+2m+|n|)}.
\end{align}
It is clear in \eqref{eq:BTZfinite} that the sum over $m,n$ reproduces the infinite volume expression, while the sum over $p$ implements periodicity in $\theta$. Explicitly we find
\be
G_E(\tau,x) = \sum_{p=-\infty}^{\infty}\frac{1}{2^\Delta(\cosh(\theta+2\pi p)-\cos\tau)^\Delta}.
\ee
Analytically continuing the Euclidean correlator to real time, we reproduce the expected singularities of the finite volume Wightman correlator starting from the Regge expansion. 

\section{Causality on the sphere}
\label{app:causalitysphere}

In this appendix we analyze the constraints of causality at finite volume. We consider the thermal retarded two-point function on the sphere,
\be
G_R(t,\theta) = \int_{- \infty}^{\infty} d \omega\, e^{- i \omega t} \sum_{{\ell}=0}^{\infty}G_{R}(\omega, \ell) \frac{\ell+\alpha}{\alpha}C_\ell^{(\alpha)}(\cos\theta) .
\ee
Let us take $0<\theta<\pi$ and set $t= \theta - \delta \theta$ with $\delta \theta>0$. Since $G(t,\theta)$ must vanish outside the lightcone, we have 
\be
\label{eq:spherecausality}
0 = G_R(\theta - \delta \theta,\theta) =  \int_{- \infty}^{\infty} d \omega\, e^{i \omega \delta \theta} \tilde G_R(\omega, \theta) ,
\ee
where 
\be
\tilde G_R(\omega, \theta) = \sum_{{\ell}=0}^{\infty} G_{R}(\omega, \ell) \frac{\ell+\alpha}{\alpha}C_\ell^{(\alpha)}(\cos\theta) e^{-i \omega \theta}.
\ee
Causality is then the statement that  $\tilde G_R(\omega, \theta)$ is analytic and sub-exponential for $\text{Im }\omega > 0$. Indeed, if this is the case
then we can close the contour in \eqref{eq:spherecausality} into the upper half-plane and get zero.

We can rewrite this condition as follows
\be\label{causality}
{\bf Causality:} ~~~ \Big|  \sum_{{\ell}=0}^{\infty}G_{R}(\omega, \ell) \frac{\ell+\alpha}{\alpha}C_\ell^{(\alpha)}(\cos\theta) \Big| \lesssim e^{- \text{Im}(\omega)\theta  } , ~~~ \text{Im }\omega  > 0 \ , 
\ee
where $0 < \theta < \pi$ and $\lesssim$ means up to subexponential corrections.

As before we can perform the Sommerfeld-Watson transform and consider the contribution of a given Regge pole 
\be
G_R(\omega , \ell) \sim {\lambda_m(\omega) \over \ell - \ell_m(\omega)}
\ee
to the left hand side of (\ref{causality}). Up to non-important power-like corrections we get 
\be
\lambda_m( \omega ) e^{- |\text{Im } \ell_m(\omega)|} \lesssim e^{- {\rm Im} (\omega)  \theta}.
\ee
This tells us that the Regge pole residues have to satisfy the following constraint
\be\label{causalitysphereRegge}
\lambda_m(\omega) \lesssim e^{- (\text{Im } \omega - | \text{Im } \ell_m(\omega) |) \pi} . 
\ee
\begin{figure}
\centering
    \includegraphics[scale=0.7]{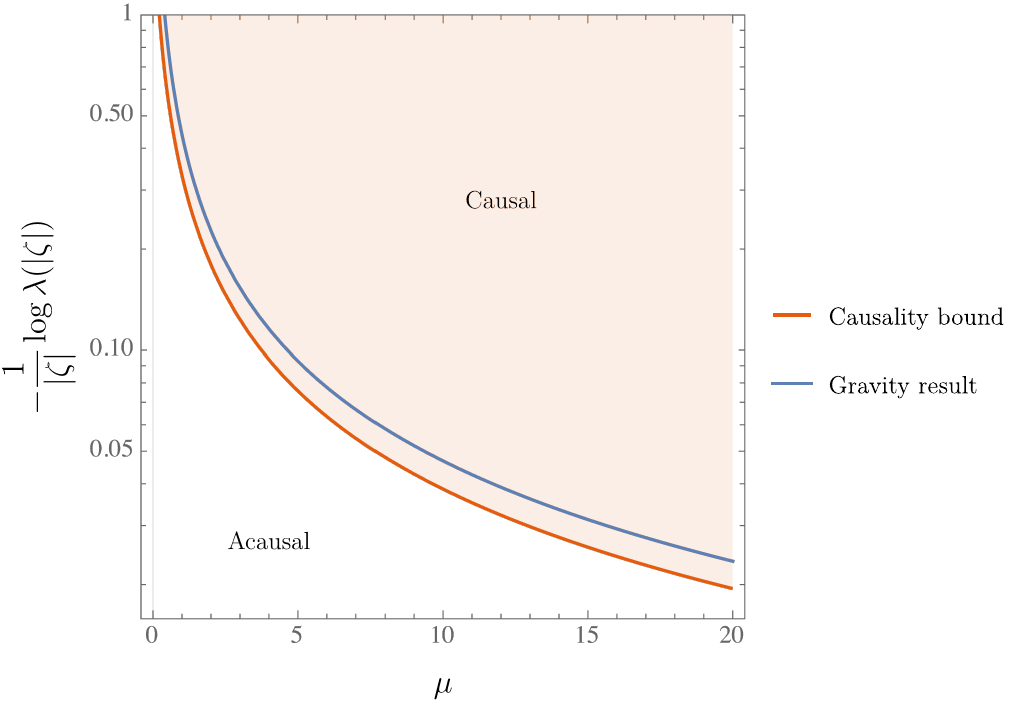}
\caption{We plot the exponential decay rate of the residues of the Regge poles \eqref{resimleikonal} as a function of $\mu$ and the corresponding causality bound \eqref{causalitysphereRegge}. As expected the gravity result is consistent with causality.}
\label{fig:causality}
\end{figure}
\indent Let us now explicitly check that \eqref{causalitysphereRegge} is satisfied, taking the case $d=4$ for simplicity. In order to compute the residues, we need to evaluate the action integral (\ref{actionintegral}) from the boundary to $z_-$,
\begin{align}\label{actionphit}
S(0,z_-)&=-\frac{p}{2}\Theta(0,z_-)+\frac{pu}{2}T(0,z_-),
\end{align}
where $\Theta(0,z_-)$ and $T(0,z_-)$ are the elapsed angle and time between 0 and $z_-$, 
\begin{align}\label{eq:TimedelayFourd}
\Theta(0,z_-)&=\frac{2r_-}{\sqrt{\mu}}K\left(\frac{r_-^2}{r_+^2}\right)\\
T(0,z_-)&=\frac{2ur_-}{\sqrt{\mu}}\frac{r_s^2\Pi\left(\frac{r_s^2}{r_+^2},\frac{r_-^2}{r_+^2}\right)+(1+r_s^2)\Pi\left(-\frac{1+r_s^2}{r_+^2},\frac{r_-^2}{r_+^2}\right)}{1+2r_s^2}.
\end{align}
Here we have defined the two turning points
\begin{align}
r_{\pm}=\sqrt{\frac{1\pm \sqrt{1-4\mu(u^2-1)}}{2(u^2-1)}}.
\end{align}
\indent Plugging the low-lying spectrum (\ref{eikonaliml}) into (\ref{actionphit}) and expanding at large $p$  gives
\begin{align}
S_m(0,z_-)&=\frac{p}{2}g_1(r_s)-\frac{m+\frac{1}{2}}{2}\log\left(\frac{g_2(r_s)ep}{m+\frac{1}{2}}\right),
\end{align}
where 
\begin{align}
g_1(r_s)&=\frac{\text{ArcCot}(\sqrt{2}r_s)}{r_s}-\frac{\text{ArcCoth}(\sqrt{2(1+r_s^2)})}{\sqrt{1+r_s^2}}\\
g _2(r_s)&=\frac{16\sqrt{2}}{(2r_s^2+1)^2}e^{\frac{\sqrt{2}}{r_s}\text{ArcCot}(\sqrt{2}r_s)}\left(\frac{2r_s^2+1}{2r_s^2+2\sqrt{2(1+r_s^2)}+3}\right)^{\frac{1}{\sqrt{2(1+r_s^2)}}}.
\end{align}
Converting the QNM residues (\ref{residuesiml}) to Regge pole residues then gives
\begin{align}\label{resimleikonal}
\underset{\hspace{2mm}\ell\to \ell_{m}(i|\zeta|)}{\text{Res }}G_R(i|\zeta|,\ell)=\frac{2\sqrt{\pi} i}{\nu\Gamma(\nu)^2m!}\left(\frac{|\zeta|\sqrt{\Omega^2-1}}{2\Omega}\right)^{2\nu}\left(\frac{|\zeta| g_2(r_s)}{\Omega}\right)^{m+\frac{1}{2}}e^{-\frac{ g_1(r_s)}{\Omega}|\zeta|}.
\end{align}
It is straightforward to check that the exponential damping factor in this expression decays fast enough to satisfy the constraint (\ref{causalitysphereRegge}), see Figure \ref{fig:causality}.

\indent One peculiar feature of the bound (\ref{causalitysphereRegge}) as opposed to the infinite volume case is that (\ref{causalitysphereRegge}) does not imply that all non-analyticities of the retarded two-point function have to satisfy $|\text{Im } \ell_m(\omega) |  > \text{Im } \omega $, and indeed the photon sphere Regge poles (\ref{eikonaliml}) explicitly violate this condition since $\Omega>1$. Instead causality allows for non-analyticities at arbitrary positions as long as the residues decay fast enough. Because of this there is no obvious generalization of the hydrohedron analysis \cite{Heller:2023jtd} to the sphere case.

\section{Late-time light cone at finite temperature in $d>2$}
\label{app:latetimelightcone}

Let us consider the following series,
\be\label{fcoft}
f_c(t) = \sum_{J=0}^{\infty} e^{i J t - c e^{-J} t} .
\ee
It mimics the sum over spinning quasi-normal modes which have exponentially small imaginary part corresponding to the probability of an orbiting particle to tunnel into a black hole, see \cite{Dodelson:2022eiz}.

The sum (\ref{fcoft}) converges in the sense of distributions, see e.g. \cite{Kravchuk:2020scc}, and gives
\be
\label{eq:zeroc}
f_0(t) = {1 \over 1 - e^{i t}} ,
\ee
which models light-cone singularities at $t= 2 \pi k$ in the vacuum, $c=0$. We would like to understand what happens to them at finite temperature, or $c\neq 0$.

To evaluate the sum for $c \neq 0$ we use the Mellin representation of the exponential
\be
e^{-z} =1+{1 \over 2 \pi i} \int_{-0 - i \infty}^{-0 + i \infty} \Gamma(s) z^{-s} d s \ .
\ee
In this way we get for the sum
\be
f_c(t) =f_0(t) + {1 \over 2 \pi i} \int_{-0 - i \infty}^{-0 + i \infty} ds \ \Gamma(s) \sum_{J=0}^\infty e^{i J t} (c e^{-J} t)^{-s}.
\ee
Here we have exchanged the sum and the integral, which is justified when both are absolutely convergent. We have $|e^{i J t} (c e^{-J} t)^{-s}| = (c t)^{- \text{Re } s} e^{J \text{Re } s}$. The sum over $J$ converges for $\text{Re } s < 0$, which explains our
treatment of the $s=0$ pole above.

Now we can trivially do the sum to get
\be
f_c(t) =f_0(t) + {1 \over 2 \pi i} \int_{-0 - i \infty}^{-0 + i \infty} ds \ \Gamma(s) {(c t)^{-s} \over 1 - e^{i t + s}} \ . 
\ee
An interesting new feature of this expression is that it has extra poles in $s$ at 
\be
s = - i t + 2 \pi i k , ~~~ k \in \mathbb{Z} \ .
\ee

There are various limits we can consider. First, let us consider $c \to 0$. In this case we close the contour to the left and we recover \eqref{eq:zeroc}. Indeed, we simply get a Taylor series in $c$.

Consider next the late-time limit $c t \gg 1$. In this case we want to deform the contour to the right. The leading contribution takes the form
\be
f_c(t) =\sum_{k=-\infty}^{+\infty} \Gamma(i (2\pi k - t)) e^{i (t- 2 \pi k) \log c t} + \mathcal{O}( e^{-e c t}),
\ee
where $\mathcal{O}( e^{-e c t})$ is the contribution of the background integral.

Notice that all the light-cone singularities and their residues stay intact, $f_c(t) \sim {i \over t - 2 \pi k}$. However, as we go away from the light-cone singularity the correlator acquires a highly oscillatory phase $e^{i (t- 2 \pi k) \log c t}$ which suppresses the correlator upon smearing. The characteristic timescale of oscillations of $e^{i {t - 2 \pi k \over \delta T}}$ is $\delta T \sim {1 \over \log c t}$, which slowly goes to zero at late times.

We expect that light-cone singularities of thermal correlators in $d>2$ behave in a similar fashion with some $\delta T(t)$ dictated by the imaginary part of the large-spin non-analyticity of the retarded two-point function, such that $\delta T(t) \to 0$ when $t \to \infty$. It would be interesting to check this explicitly.

\section{Deriving the representation \eqref{eq:gensumformula}}
\label{app:derivingwindings}
Our starting point is the sum over Regge poles \eqref{reggefinitevolume},
\be \label{reggefinitevolumeApp}
G_E(\tau,\theta) = -\frac{\pi}{\alpha\beta}\sum_{n=-\infty}^{\infty}\sum_m e^{i \zeta_n \tau}\frac{k_{mn} C_{k_{mn}-\alpha}^{(\alpha)}(-z)}{\sin\left(\pi(k_{mn}-\alpha)\right)} \underset{\hspace{2mm}k\to k_{mn}}{\text{Res}}G_{R}\left(i |\zeta_n|,k-\alpha\right) . \nn
\ee 
We can use the following representation for Gegenbauer polynomials valid for $0 < \theta < \pi$
\be
C_{k-\alpha}^{(\alpha)}(-z) = \frac{i (2\sin\theta) ^{1-2 \alpha } \Gamma
   (k+\alpha )}{\Gamma (\alpha ) \Gamma (k+1)} \Big( e^{- i (1+k-\alpha)(\pi - \theta)}f(k,\theta) -e^{ i (1+k-\alpha)(\pi - \theta)}f(k,\pi - \theta)\Big),
\ee
where
\be
f(k,\theta) = \ _2 F_1 (1-\alpha, 1+k-\alpha, 1+k, e^{2 i \theta}). 
\ee
Since the sum over Regge poles runs over $\text{Im }k>0$, we can also expand
\be
{1 \over \sin \pi (k- \alpha)} =- 2 i \sum_{j=0}^{\infty} e^{i \pi (2j+1) (k-\alpha)}.
\ee

In this way we get the following representation for the correlator
\be
\label{eq:sumim}
G_E(\tau, \theta) = \sum_{j=0}^\infty \Big( g_E(\tau, \theta+ 2 \pi j ) + (-1)^{2\alpha}g_E(\tau, 2 \pi - \theta+ 2 \pi j ) \Big) \ ,
\ee
where
\be
g_E(\tau,\theta) &=  {4^{1-\alpha} \over (\sin \theta)^{2 \alpha-1}}\frac{\pi}{\alpha\beta}\sum_{n=-\infty}^{\infty} e^{i \zeta_n \tau} \sum_m \underset{\hspace{2mm}k\to k_{mn}}{\text{Res}}G_{R}\left(i |\zeta_n|, k-\alpha\right) e^{i (1+k_{mn}-\alpha) \theta} \nn \\ 
&\hspace{5 mm}\times {\Gamma(k_{mn}+\alpha) \over \Gamma(k_{mn})\Gamma(\alpha)} \ _2 F_1 (1-\alpha,1+k_{mn}-\alpha,1+k_{mn}, e^{2 i \theta}) \ . 
\ee
In writing the formula above we implicitly used that $f(k, \theta) = f(k, \theta+2 \pi j)$. Given that $\theta=0$ is a branch point of the hypergeometric function, periodicity of  $f(k, \theta)$ in $\theta$ is only true given a particular prescription for going around this branch point, namely $f(k, \theta+i \epsilon)$. This prescription is understood in \eqref{eq:sumim}.

\section{Astrophysical black holes}
\label{sec:Signatures}

Astrophysical black holes are typically described by the Kerr family of asymptotically flat metrics, which is parameterized by the mass $M$ and angular momentum $J$ of the hole.
Following the 2019 release by the Event Horizon Telescope (EHT) of the first image of a supermassive black hole in our sky \cite{EventHorizonTelescope:2019dse}, it was soon realized that such black hole images ought to display a ``photon ring'' consisting of multiple mirror images of the main emission surrounding the hole: this is a generic prediction of general relativity, which follows directly from the observation that a Kerr black hole possesses a ``photon shell'' of (unstably) bound photon orbits outside its event horizon, in which light can orbit the black hole (possibly multiple times) before escaping to a distant observer  \cite{Gralla:2019xty,Johnson:2019ljv,Gralla:2019drh}.
Whilst this theoretically predicted feature has not yet been resolved with the ground-based EHT, detecting the photon ring will be a key target for future spaceborne interferometric observations of the supermassive black holes M87* and Sgr\,A* \cite{Gralla:2020srx,Gurvits:2022wgm,Cardenas-Avendano:2022csp,Hudson:2023tuy}.

\subsection{Observational signatures of a Schwarzschild black hole}

Black holes predict a very specific shape and structure for the photon ring, which we will now briefly describe, and which can be used to answer the question: ``How can one ascertain whether an astrophysical source is truly a \textit{Kerr} black hole, as predicted by the Kerr hypothesis in general relativity?''

\begin{figure}
    \centering
    \includegraphics[width=0.32\textwidth]{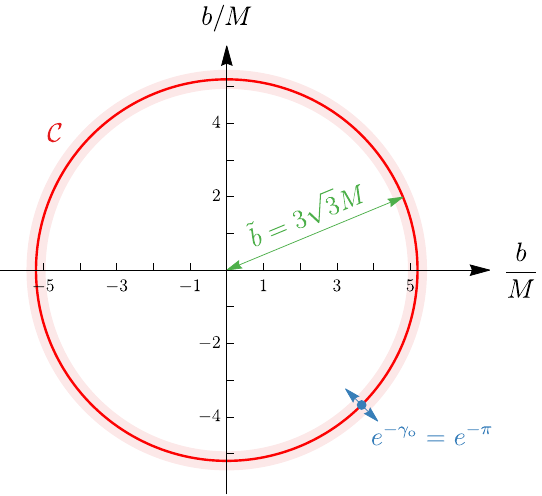}
    \includegraphics[width=0.32\textwidth]{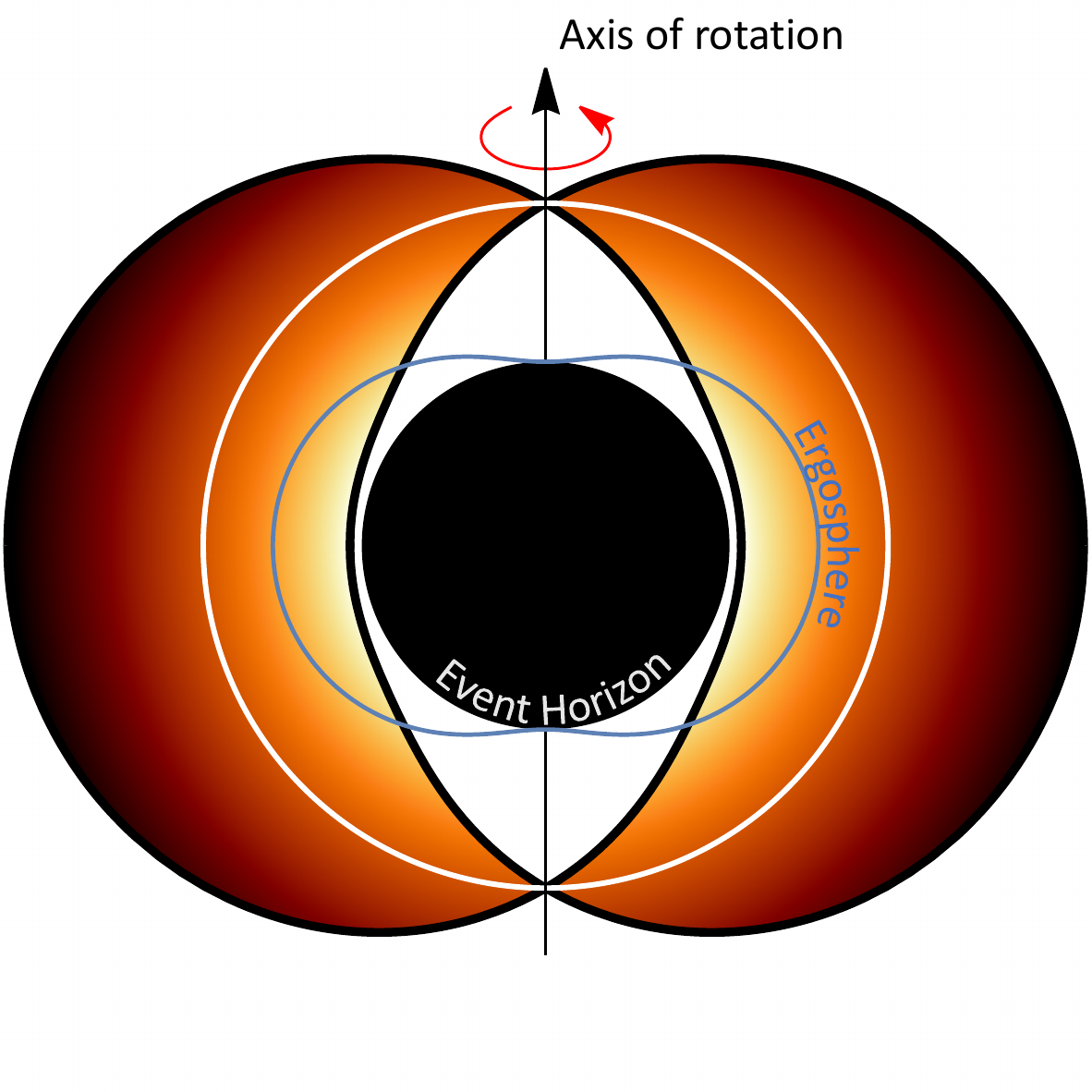}
    \includegraphics[width=0.32\textwidth]{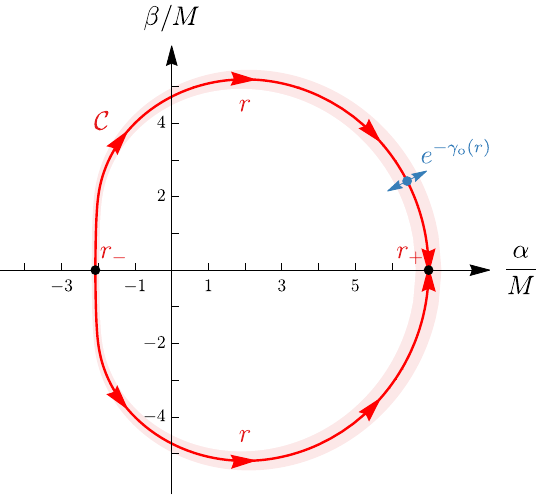}
    \caption{\textbf{Left:} Image plane of a distant observer in the Schwarzschild spacetime.
    A photon with total angular momentum $L$ and energy $E$ appears at impact parameter $b=L/E$.
    Photons whose energy-rescaled angular momentum takes the critical value $\tilde{b}=3\sqrt{3}M$ can be trapped on the photon sphere of unstably bound light orbits at $r=3M$, whose image is the ``critical curve'' $\mathcal{C}$ (red).
    \textbf{Middle:} When a black hole spins, its photon sphere thickens into a shell bounded by circular-equatorial photon orbits, with the innermost prograde orbit at $r=r_+$ and the outermost retrograde orbit at $r=r_-$.
    As the black hole spins up to extremality ($|J|\to M^2$), the shell reaches its maximal size $(r_+,r_-)\to(M,4M)$.
    In Kerr, only one orbital radius $r=r_0\in(r_+,r_-)$, which corresponds to zero-angular-momentum orbits with $p_\phi=0$, allows for light to pass over the poles (white photon sphere, located at $r=3M$ in Schwarzschild).
    Bound orbits inside/outside the sphere are prograde/retrograde with $p_\phi\gtrless0$.
    \textbf{Right:} Image plane of a distant, equatorial observer ($\theta_{\rm o}=\pi/2$) in the Kerr spacetime with spin $J/M^2=99.9\%$, parameterized by Bardeen's Cartesian coordinates $(\alpha,\beta)$ \cite{Bardeen1973}.
    The critical curve $\mathcal{C}$ (red) is now parameterized by the photon shell radius $r$ that a light ray traced backwards into the geometry ends up orbiting at.
    The Lyapunov exponent $\gamma(r)$ governing the instability of the bound orbit at radius $r$ also controls the width of the photon ring (red annulus) at the corresponding angle around the critical curve (for the non-rotating black hole, $\gamma=\pi$).
    Measuring the size, shape, and radial profile of the photon ring can yield information on the parameters of the black hole.}
    \label{fig:PhotonRing}
\end{figure}

We begin with a discussion of the non-rotating Schwarzschild black hole of mass $M$, both due to its relative simplicity and because it can be recovered from its AdS counterpart---the subject of this paper---in the limit $R_{\text{AdS}}/M\to\infty$ (see Section \ref{sec:flatspace}).
Consider an observer at a large distance $D\gg M$ from the black hole in asymptotically flat spacetime.
A photon received with energy $E=-p_t$ and total angular momentum $L^2=p_\theta^2+p_\phi^2\csc^2{\theta}$ appears in the image plane of the observer at impact parameter $b=L/E$.
As is now well-known, a photon shot back from the observer with critical energy-rescaled angular momentum $\tilde{b}=3\sqrt{3}M$ will asymptote to an unstable planar orbit on the photon sphere at $r=3M$.\footnote{Remarkably, the critical image radius $\tilde{b}=3\sqrt{3}M$ was already published (by David Hilbert) in 1917, immediately following Einstein's publication of his general theory of relativity in 1915, and mere months after Schwarzschild found his eponymous solution to Einstein's field equations in 1916.}
Photons traced backwards into the geometry with $b<\tilde{b}$ must fall into the black hole, while those with $b>\tilde{b}$ are merely deflected and eventually escape back out to infinity.
Thus, the image $b=\tilde{b}$ of the photon sphere---which is also known as the ``critical curve'' and is depicted in red in Figure \ref{fig:PhotonRing} (left panel)---is the boundary delineating the region of photon capture from that of photon escape.
Photons shot back from the vicinity of the critical curve (i.e., with small $\delta b=b-\tilde{b}$) will describe multiple orbits skirting the photon sphere before eventually falling into the black hole (if $\delta b<0$) or escaping back out to infinity (if $\delta b>0$).
The number $n$ of half-orbits executed around the black hole diverges logarithmically in the perpendicular distance from the curve \cite{Gralla:2019drh,Johnson:2019ljv},
\begin{align}
    \label{eq:LogDivergence}
    n\approx-\frac{1}{\gamma_{\rm o}}\log\left|\frac{\delta b}{\tilde{b}}\right|,
\end{align}
where $\gamma_{\rm o}=\pi$ is the Lyapunov exponent governing the orbital instability of nearly bound rays near the Schwarzschild photon sphere.
The upshot is that a light ray appearing at a distance $|\delta b|$ from the critical curve must be aimed exponentially closer to it by $e^{-\gamma_{\rm o}}=e^{-\pi}\approx4.3\%$ in order to execute an additional half-orbit around the black hole.\footnote{The $\delta b>0$ version of these statements was already known to Luminet in 1979 \cite{Luminet:1979}.}

In particular, if a black hole is surrounded by an astrophysical source---such as an equatorial disk of emission, for instance---then multiple images of the source will appear in the vicinity of the critical curve \cite{Johnson:2019ljv}.
More precisely, there could in principle be a whole infinite sequence of images of the main emission, with each successive one appearing smaller (and closer to the critical curve) than its predecessor by a factor of $e^{-\gamma_{\rm o}}$.
These images would thus be lensed into a ``photon ring'' with an intricate substructure consisting of self-similar subrings stacked on top of one another, and with relative widths decreasing by a demagnification factor of $e^{-\gamma_{\rm o}}$.

In the idealized configuration (first studied by Falcke et al. \cite{Falcke:1999pj} and then revisited by Narayan et al. \cite{Narayan:2019imo}) of a black hole that is fully immersed within a spherically symmetric accretion flow of hot, radiating gas, the event horizon casts a ``shadow'' on the surrounding emission, and the black hole image displays a darkness depression whose edge precisely coincides with the critical curve.
Moreover, in that case, the observed intensity near the edge of the shadow scales like $I\propto n$, and therefore diverges logarithmically according to \eqref{eq:LogDivergence} (though, in practice, absorption effects cut off the divergence after some finite number of orbits).
As a result, the critical impact parameter $\tilde{b}$ is precisely the radius of the observed shadow, and the Lyapunov exponent can in principle be read off from the intensity profile near its edge.
In the currently favored scenario of near-equatorial emission \cite{Chael:2021rjo}, the image decomposes into discrete subrings labeled by half-orbit index $n$, whose asymptotic radius as $n\to\infty$ again approaches $\tilde{b}$, and whose relative widths scale with a factor of $w_{n+1}/w_n\stackrel{n\gg1}{\approx}e^{-\gamma_{\rm o}}$, such that the critical parameters $\tilde{b}$ and $\gamma_{\rm o}$ are both still measurable.

Although no photon ring has been measured so far, this GR-predicted structure is in principle observable; moreover, it is universal (i.e., very weakly dependent on the nature of the source), as it follows purely from the lensing behavior of the black hole, which is in turn determined by the geometry of its spacetime (in particular, of its photon sphere).
State-of-the-art simulations numerically confirm this behavior \cite{Johnson:2019ljv,Chael:2021rjo}, so future observations may well measure the photon sphere critical parameters.

Finally, the photon sphere also controls the spectrum $\omega_{\ell mn}^{(s)}$ of quasi-normal modes in the eikonal regime of high-frequency $\omega\gg1/M$.
In that limit, one can apply the geometric-optics approximation and use congruences of null geodesics to describe massless waves (of any spin $s$, whose effect is subleading).
Moreover, it turns out that the quasi-normal mode boundary conditions correspond to light rays that are asymptotically bound in the photon sphere, so it should come as no surprise that the eikonal QNM spectrum is controlled by the critical parameters of the photon sphere \cite{Goebel:1972,Ferrari:1984zz,Mashhoon:1985,Schutz:1985km,Iyer:1986np,Berti:2009kk},
\begin{align} 
    \label{eq:SchQNM}
    \omega_{\ell mn}^{(s)}\stackrel{\ell\gg|s|}{\approx}\left(\ell+\frac{1}{2}\right)\Omega-i\left(n+\frac{1}{2}\right)\gamma,\qquad
    \Omega=\gamma=\frac{1}{\tilde{b}}=\frac{1}{3\sqrt{3}M},
\end{align}
where $\Omega$ is the angular velocity of bound photon orbits, while the Lyapunov-exponent-in-time $\gamma\approx dr/dt$ is equal to the Lyapunov-exponent-per-orbit $\gamma_{\rm o}=\pi\approx dr/dn$, divided by the time elapsed per half-orbit, $\tau_{\rm o}\equiv(\Delta t)_\text{half-orbit}=3\sqrt{3}\pi M\approx dt/dn$.

\subsection{Observational signatures of a Kerr black hole}

We continue our discussion of how to ascertain whether an astrophysical source
is truly a black hole, now extending it to the general case of a rotating (Kerr) black hole.
Here we sketch only the main ideas, which are essentially the same as in the previous section, and omit most explicit formulas.

In the presence of rotation, the photon sphere thickens into a shell (Figure \ref{fig:PhotonRing}, middle panel) containing multiple orbital radii $r_+<r<r_-$, where $r_\pm$ denote the radii of the prograde/retrograde circular-equatorial orbits bounding the shell.
Only light trapped at the radius $r_0\in(r_+,r_-)$ of the zero-angular-momentum orbits can pass over the poles.
The sphere $r=r_0$ generalizes the Schwarzschild photon sphere, and indeed $r_0\to3M$ as $J\to0$.
Orbits with radius $r\in(r_+,r_0)$ within this photon sphere all have positive angular momentum and are corotating with the black hole (prograde), while those with radius $r\in(r_0,r_-)$ outside the photon sphere all have negative angular momentum and are counter-rotating relative to the black hole (retrograde).
These orbits describe librations (polar oscillations) up to some angle $\tilde{\theta}(r)$, which equals $\pi/2$ at the edges $r_\pm$ of the shell and vanishes at $r_0$.
One may assign a unique ``signed inclination'' $\mu_{\rm o}(r)=\pm\sin{\tilde{\theta}(r)}$ to each orbit, where the sign $\pm$ is that of the orbital angular momentum: thus, $\mu_{\rm o}(r)$ decreases monotonically from $\mu_{\rm o}(r_+)=1$ through $\mu_{\rm o}(r_0)=0$ to $\mu_{\rm o}(r_-)=-1$, resulting in a 1-1 correspondence between photon shell radii $r\in[r_+,r_-]$ and signed inclinations $\mu_{\rm o}\in[-1,1]$.

In the absence of spherical symmetry, the lensing behavior of the black hole depends on the polar inclination $\theta_{\rm o}$ of the distant observer relative to the spin axis.
Following Bardeen \cite{Bardeen1973}, it is convenient to parameterize the image plane of the observer using Cartesian coordinates $(\alpha,\beta)$ defined such that a photon received with four-momentum $p^\mu$ appears at a position $\alpha=-p_\phi/(p_t\sin{\theta_{\rm o}})$ and $\beta=-p_\theta/p_t$.
The critical curve is then defined as the image of the photon shell, that is, the set of image-plane directions corresponding to light rays that are asymptotically bound in the photon shell.
Bardeen \cite{Bardeen1973} provides a simple analytic expression for this curve.
The key difference with the non-rotating case is that the curve is now parameterized by photon shell radius, with each point $(\tilde{\alpha}(r),\tilde{\beta}(r))$ on (half of) the critical curve corresponding to a light ray that is trapped at a different orbital radius $r$ in the photon shell (Figure \ref{fig:PhotonRing}, right panel).

The (energy-rescaled) azimuthal angular momentum $\lambda=-p_\phi/p_t$ of a critical ray with horizontal impact parameter $\tilde{\alpha}(r)$ takes the value $\lambda_{\rm o}(r)=-\tilde{\alpha}(r)\sin{\theta_{\rm o}}$, which is thus a measurable quantity.
Each bound photon orbit has a different rate of orbital instability governed by its own Lyapunov exponent $\gamma_{\rm o}(r)$, and a different half-orbit time lapse $\tau_{\rm o}(r)=(\Delta t)_\text{half-orbit}$.
Explicit forms for these critical parameters of the Kerr black hole are given in terms of elliptic integrals in \cite{Johnson:2019ljv,Gralla:2019drh}.

The half-orbit number still diverges logarithmically as one approaches the critical curve, so \eqref{eq:LogDivergence} still holds, except that $\gamma_{\rm o}(r)$ now varies around the curve.
Still, each Lyapunov exponent $\gamma_{\rm o}(r)$ can in principle be read off from the radial intensity profile near the corresponding position $(\tilde{\alpha}(r),\tilde{\beta}(r))$ around the photon ring: depending on the precise astrophysical scenario, this could be done either by comparing the angle-dependent demagnification $e^{-\gamma_{\rm o}(r)}$ across successive subrings (when the emission is near-equatorial), or from the log-divergence in the intensity (when the emission is spherical).
Likewise, $\lambda_{\rm o}(r)$ can in principle be read off from the horizontal impact parameter of light in the photon ring, so both $\lambda_{\rm o}(r)$ and $\gamma_{\rm o}(r)$ are observable, as is $\tau_{\rm o}(r)$, which can for instance be inferred from the time delay of light echoes \cite{Hadar:2020fda}.
Thus, precise measurements of the photon ring (its GR-predicted substructure and critical exponents) could in principle be used to test whether a source is indeed a Kerr black hole.

Finally, one expects the eikonal QNM spectrum to still be described by the Kerr photon shell, as in the Schwarzschild case.
Though this was indeed a widely held picture since the seminal work of Ferrari \& Mashhoon \cite{Ferrari:1984zz} and Iyer \& Wald \cite{Iyer:1986np} in the 1980s, their work did not in fact apply to Kerr, as the class of potentials they considered did not include the Kerr radial geodesic potential.
Thus, this lore was only verified to hold in 2012 \cite{Yang:2012he}, using indirect formulas for the form of the spectrum.
Following the explicit derivation of photon shell critical exponents, the analogue of \eqref{eq:SchQNM} was finally derived last year \cite{Hadar:2022xag}: as $\ell,m\to\infty$ with $\tilde{\mu}=m/\ell$ held fixed,
\begin{align}
    \omega_{\ell mn}^{(s)}\stackrel{\ell\gg|s|}{\approx}\left(\ell+\frac{1}{2}\right)\Omega(\tilde{\mu})-i\left(n+\frac{1}{2}\right)\gamma(\tilde{\mu}),\quad
    \Omega(\tilde{\mu})=\frac{\tilde{\mu}}{\lambda_{\rm o}(r)},\quad
    \gamma(\tilde{\mu})=\frac{\gamma_{\rm o}(r)}{\tau_{\rm o}(r)},
\end{align}
where $\tilde{\mu}=m/\ell\in[-1,1]$ and the photon shell radius $r\in[r_+,r_-]$ are bijectively related by identifying $\tilde{\mu}(r)\equiv\mu_{\rm o}(r)$ with the signed inclination $\mu_{\rm o}(r)=\pm\sin{\tilde{\theta}(r)}$ defined above.

\bibliographystyle{JHEP}
\bibliography{main}
  
\end{document}